\documentclass[aps,showpacs,twocolumn,
superscriptaddress]{revtex4}
\usepackage{graphicx}% Include figure files
\usepackage{dcolumn}% Align table columns on decimal point
\usepackage{bm}% bold math
\usepackage{color}
\usepackage[normalem]{ulem} % \sout{old text} for strikeout
\usepackage[dvipsnames]{xcolor} % For blue in-text comments
\usepackage{hyperref}
\hypersetup{
  colorlinks=true,        % false: boxed links; true: colored links
  linkcolor=blue,         % color of internal links
  citecolor=cyan,         % color of links to bibliography
}
\usepackage{mathrsfs}
\usepackage[utf8]{inputenc}
\usepackage{mathtools}
\usepackage{doi}
\usepackage{amsmath}
\usepackage{amssymb}

\begin{document}
	
%%%
\title{Dynamics of charged particles and magnetic dipoles around magnetized quasi-Schwarzschild black holes}
%%%%

\author{Bakhtiyor Narzilloev}
\email[]{nbakhtiyor18@fudan.edu.cn}
\affiliation{Center for Field Theory and Particle Physics and Department of Physics, Fudan University, 200438 Shanghai, China }
\affiliation{Akfa University, Kichik Halqa Yuli Street 17, Tashkent 100095, Uzbekistan}
\affiliation{Ulugh Beg Astronomical Institute, Astronomy Str. 33, Tashkent 100052, Uzbekistan}
\author{Javlon Rayimbaev}
\email[]{javlon@astrin.uz}
\affiliation{Ulugh Beg Astronomical Institute, Astronomy Str. 33, Tashkent 100052, Uzbekistan}
\affiliation{Akfa University, Kichik Halqa Yuli Street 17,  Tashkent 100095, Uzbekistan}
\affiliation{Institute of Nuclear Physics, Ulugbek 1, Tashkent 100214, Uzbekistan}
\affiliation{National University of Uzbekistan, Tashkent 100174, Uzbekistan}
\author{Ahmadjon~Abdujabbarov}
\email[]{ahmadjon@astrin.uz}
\affiliation{Shanghai Astronomical Observatory, 80 Nandan Road, Shanghai 200030, P. R. China}
\affiliation{Ulugh Beg Astronomical Institute, Astronomy Str. 33, Tashkent 100052, Uzbekistan}
\affiliation{Institute of Nuclear Physics, Ulugbek 1, Tashkent 100214, Uzbekistan}
\affiliation{National University of Uzbekistan, Tashkent 100174, Uzbekistan}
\affiliation{Tashkent Institute of Irrigation and Agricultural Mechanization Engineers, Kori Niyoziy, 39, Tashkent 100000, Uzbekistan}
\author{Bobomurat Ahmedov}
\email[]{ahmedov@astrin.uz}
\affiliation{Ulugh Beg Astronomical Institute, Astronomy Str. 33, Tashkent 100052, Uzbekistan}
\affiliation{Tashkent Institute of Irrigation and Agricultural Mechanization Engineers, Kori Niyoziy, 39, Tashkent 100000, Uzbekistan}
\affiliation{National University of Uzbekistan, Tashkent 100174, Uzbekistan}
\author{Cosimo Bambi}
\email[]{bambi@fudan.edu.cn}
\affiliation{Center for Field Theory and Particle Physics and Department of Physics, Fudan University, 200438 Shanghai, China }

\date{\today}

\begin{abstract}

In the present paper, we have investigated the motion of charged particles together with magnetic dipoles to determine how well the spacetime deviation parameter $\epsilon$ and external uniform magnetic field can mimic the spin of a rotating Kerr black hole. Investigation of charged particle motion has shown that the deviation parameter $\epsilon$ in the absence of external magnetic fields can mimic the rotation parameter of Kerr spacetime up to $a/M \approx0.5$. The combination of external magnetic field and deviation parameter can do even a better job mimicking the rotation parameter up to $a/M\simeq0.93$, which corresponds to the rapidly rotating case. Study of the dynamics of magnetic dipoles around quasi-Schwarzschild black holes in the external magnetic field has shown that there are degeneracy values of ISCO radius of test particles at $\epsilon_{cr}>\epsilon\geq 0.35$ which may lead to two different values of the innermost stable circular orbit (ISCO) radius. When the deviation parameter is in the range of $\epsilon \in (-1,\ 1)$, it can mimic the spin of a rotating Kerr black hole in the range $a/M \in (0.0537, \ 0.3952)$ for magnetic dipoles with values of magnetic coupling parameter $\beta \in [-0.25,\ 0.25]$ in corotating orbits. 
 
\end{abstract}


\maketitle

\section{Introduction}

\textcolor{black}{The first exact analytical solution of the vacuum field equations of Einstein’s general relativity has been obtained just
after its discovery in 1916 by Schwarzschild~\cite{Schwarzschild1916} and describes exterior spacetime of the non-rotating spherically symmetric  black hole.} The rotating black hole solution is obtained by Kerr and includes two parameters: the total mass of the black hole and its rotation parameter. Most observational features of the astrophysical black holes can be, in principle, explained by the solution describing the Kerr black hole. On the other hand one may alternatively consider the extension of the Kerr solution with additional parameters,~see, for example,~\cite{Newman63,Zimmerman89,Glampedakis06b,Johannsen10,Johannsen11,Johannsen13,Konoplya16,Cardoso14,Sen92,Rezzolla2014}. 
The electric charge may also affect the gravitational field of the charged black hole and the properties of such objects have been studied in~\cite{Grunau11,Zakharov94,Stuchlik02,Pugliese10,Pugliese11b,Pugliese11,Patil12} for different astrophysical scenarios. Black hole may be considered as embedded on the brane of higher dimensional spacetime, see the Refs.~\cite{Turimov17,Whisker05,Majumdar05,Liang17,Li15} where several properties of back holes with brane charge have been studied. 
Black holes may have gravitomagnetic monopole charge and authors of Refs.~\cite{Liu11,Zimmerman89, Morozova09, Aliev08, Ahmedov12, Abdujabbarov11, Abdujabbarov08} have studied the properties of spacetime with nonvanishing gravitmagnetic charge.
Authors of Refs.~\cite{Bambi17c,Rayimbaev15, Bambi11b, Chen12, Narzilloev:2020b, Bambi12,Bambi13e,Bambi17b,Cao18} have studied deformed spacetime of black holes and its properties. 

One of the interesting extension of the Kerr solution has been proposed in~\cite{Glampedakis06b} where an approximate solution of Einstein vacuum equations has been obtained. It was also introduced the leading order deviation from Kerr solution due to spacetime quadrupole moment. The spacetime properties around so-called quasi-Kerr black hole have been studied in Refs.~\cite{Psaltis12, Liu12b}. In a previous paper, we have studied the weak lensing near the quasi-Kerr black hole~\cite{Chakrabarty18}. Recently we have also studied the charged particle motion around quasi-Kerr compact object in the presence of magnetic field~\cite{Narzilloev19}.  

Testing general relativity and alternative theories of gravity through gravitational lensing and motion of test particles in the various conditions is important to distinguish the central black hole parameters from the ones of alternate gravity theories, since their effects are similar or exactly the same at some range of values of the parameters of  theories of gravity. For example in our previous works we have shown how the effects of MOG field parameters~\cite{Haydarov20b}, conformal parameters\cite{Haydarov20,Narzilloev20c}, the coupling parameter of the Einstein-Gauss-Bonnet theory~\cite{Abdujabbarov2020PDU}, electric charge of a black hole in Einstein-Maxwell theory \cite{TurimovPhysRevD2020}, stringy charge \cite{Narzilloev20c}, perfect fluid dark matter \cite{Narzilloev20b} and quantum gravity \cite{Rayimbaev20d} can mimic the spin of rotating Kerr black hole. On the other hand, by now, in spite of the attempts to detect neutron stars as recycled radio pulsars near the supermassive black hole Sagittarius A* in the center of Milky way galaxy by GRAVITY collaboration, we do not have any astrophysical observations of them. One of the reason of the absence of the pulsars around SgrA* is scattering of radio wave in the plasma medium surrounding the SMBH and the other one might be the dominated effects of the magnetic interaction between the neutron star's dipole moment and the magnetic field around the black hole created by either magnetic charge of the central black hole or electric current of accreting matter .
The stable circular and chaotic motions of neutral particles \cite{Radim2019EPJC}, dynamics and quasihormonic oscillations of charged particles around static and rotating black holes immersed in external asypmtotically uniform magnetic fields ~\cite{Kolos15,Tursunov16,Stuchlik16,Stuchlik20,Tursunov2020ApJ} and plasma magnetosphere surrounding black holes in different gravity models have been analyzed in detail by the authors of Refs.\cite{Takahashi2009ApJ,Kopacek2010ApJ,Kopacek2014ApJ,LiDan2019EPJP,YiMiao2020PhyS} in particular using the method of the Lyapunov to show the difference between regular and chaotic orbits. It is also shown that  even a small misalignment and frame dragging effects cause to reduce the chaotic motion.

Recent observation of image of supermassive black hole at the center of elliptical galaxy M87~\cite{EHT19a,EHT19b} and detection of the gravitational waves by LIGO-Virgo collaboration~\cite{LIGO16a,LIGO16} provided the test of general relativity in strong field regime. 
In fact gravitational wave generated by the binary compact objects and the wave properties strongly depend on their spin and chaos degree in the system. The existence of chaotic motion in such systems in the  problems of two spinning black holes in the post-Newtonian approximation has been shown in Refs.~\cite{Levin2000PhRvL,Levin2003PhRvD,Cornish2002PhRvL,Cornish2003PhRvD,Buonanno2005PhRvD,Wu2007PhRvD,Wu2008PhRvD,Wu2010PhRvD,Wu2015PhRvD,Wu2015MNRAS,Zhong2010PhRvD82,Mei2013EPJC}. 

At the same time these experiments and observations open window for testing the modified and alternative theories of gravity together with analysis of X-ray observations from active galactic nuclei (AGN) ~\cite{Bambi16c,Zhou18,Tripathi19}. The second generation Very long baseline interferometer (VLBI) instrument GRAVITY through precise observations of highly relativistic motions of matter and S2 star close to Sgr A* has also provided experimental tests of general relativity in strong field. 

Despite the fact that in general relativity the black hole cannot have its own magnetic field due to no-hair theorem~\cite{Misner73}, the latter can be considered as immersed in external magnetic field~\cite{Wald74} created by current of electric charges in accretion disk. The spacetime curvature will change the original structure of the external magnetic field. The detailed specifications of these changes and test particle motion around compact object in the presence of magnetic field have been studied in Refs.~ \cite{Chen16,Hashimoto17,Dalui19,Han08,Moura00,Morozova14,Jawad16,Hussain15,Jamil15,Hussain17,Babar16,Banados09,Majeed17,Zakria15,Brevik19,DeLaurentis2018PhRvD, Shaymatov13,Atamurotov13a,Narzilloev20a,Narzilloev20b}.  
The structure of the electromagnetic field around compact objects in alternate and modified theories of gravity have been explored in Refs.~\cite{Kolos17,Kovar10,Kovar14,Aliev89,Aliev02,Aliev86,Frolov11,Frolov12,Stuchlik14a,Shaymatov14,Abdujabbarov10,Abdujabbarov11a,Karas12a,Shaymatov15,Stuchlik16,Rayimbaev20,Turimov18b,Shaymatov20egb,Shaymatov19b,Rayimbaev19,Rayimbaev20b,Rayimbaev19b,Shaymatov20b,Narzilloev20c}. Quantum interference effects in conformal Weyl gravity has been studied in~\cite{Hakimov17}. Periodic circular orbits, regular orbits and chaotic orbits of neutral and charged particles around various black holes have been investigated in Refs.\cite{Karas1992GReGr,Nakamura1993ApSS,Kopacek2010ApJ,Kopacek2014ApJ,LiDan2019EPJP,YiMiao2020PhyS,Takahashi2009ApJ}.
The magnetic dipole motion around black hole in the presence of asymptotically uniform magnetic field has been studied in~\cite{deFelice,defelice2004} with further development to the case of modified gravity theories in~\cite{Rayimbaev16,Oteev16,Toshmatov15d,Abdujabbarov14,Rahimov11a,Rahimov11,Narzilloev20c,Haydarov20,Haydarov20b,Rayimbaev20d,Vrba20,Abdujabbarov20}.

In this work our main purpose is to study the charged particles and magnetic dipoles motion around magnetized quasi-Schwarzschild black hole. The paper is organized as follows: Sect.~\ref{sectt2} is devoted to study the dynamics of charged particles around quasi-Schwarzschild black hole and comparison with one in Kerr spacetime. The magnetic dipoles motion around quasi-Schwarzschild black hole immersed in an external magnetic field is explored in Sect.~\ref{sectt3}. In this section the obtained results have been compared with particles dynamics around Kerr black hole. Then the obtained results have been applied to real astrophysical scenarios in Sect.~\ref{sectt4}. We conclude our results in Sect.~\ref{conclusion}. Throughout the paper we use the spacelike signature (-,+,+,+) and system of units where $G=1=c$. 

\section{Charged particle motion. Quasi-Schwarzschild versus Kerr black hole \label{sectt2}}

\subsection{Magnetic field around compact object}

Before going through the investigation of magnetized particles motion around quasi-Schwarzschild compact object immersed in an external asymptotically uniform magnetic field we start with the case when the particle is electrically charged only. 
	The quasi-Schwarzschild spacetime metric can be obtained from rotating quasi-Kerr one~\cite{Glampedakis06b} using the following decomposition $$g_{\mu\nu}=g_{\mu\nu}^{Schw}+\epsilon h_{\mu\nu} ,$$ where $g_{\mu\nu}^{Schw}$ corresponds to the standard Schwarzschild metric and $\epsilon h_{\mu\nu}$ corresponds to the deviation from the Schwarzschild spacetime. Parameter $\epsilon$ defines the deviation from the spherically symmetric spacetime due to the additional term in mass quadrupole moment $Q$ of the gravitating object as $$Q=-\epsilon M^3\ ,$$ and it can take both negative and positive signs~\cite{Glampedakis06b}.
	In the linear approximation in $\epsilon$ the contravariant components of the spacetime metric can be written as $$g^{\mu\nu}=g^{\mu\nu}_{Schw}-\epsilon h^{\mu\nu} ,$$ and thus the upper indices of $h^{\mu\nu}$ can be lowered by using Schwarzschild metric tensor. The contravariant components of $h^{\mu\nu}$ are given by the expressions (see~\cite{Glampedakis06b})
	\begin{eqnarray}\nonumber
	h^{tt}&=&f^{-1} (1-3 \cos^2\theta) F_1 ,\\\nonumber
	h^{rr}&=&f (1-3 \cos^2\theta) F_1 ,\\\nonumber
	h^{\theta\theta}&=&-\frac{1}{r^2} (1-3 \cos^2\theta) F_2 ,\\\nonumber
	h^{\phi\phi}&=&-\frac{1}{r^2 \sin^2\theta} (1-3 \cos^2\theta) F_2 ,\\\nonumber
	f&=&1-\frac{2 M}{r} ,
	\end{eqnarray}
		where radial functions $F_1$ and $F_2$ read
	\begin{eqnarray}
	F_1&=&-\frac{5 (r-M) \left(2 M^2+6 M r-3 r^2\right)}{8 M r (r-2 M)}\\\nonumber
	&&-\frac{15 r (r-2 M) \log
		\left(\frac{r}{r-2 M}\right)}{16 M^2}\ ,
	\\
	F_2&=&\frac{5 \left(2 M^2-3 M r-3 r^2\right)}{8 M r}
	\\\nonumber
	&&+\frac{15 \left(r^2-2 M^2\right) \log
		\left(\frac{r}{r-2 M}\right)}{16 M^2}\ .
	\end{eqnarray}
After lowering the indices of $h^{\mu\nu}$ with the use of $g_{\mu\nu}^{Schw}$ the quasi-Schwarzschild spacetime metric takes the following form
	\begin{eqnarray}\label{metric}
	ds^2=g_{tt}dt^2
	+ g_{rr}dr^2
	+ k(r,\theta) r^2 d\Omega^2\ , 
	\end{eqnarray}
	where
	\begin{eqnarray}
	g_{tt}&=&-f \left[1-\epsilon F_1 (1-3 \cos^2\theta)\right], \\
	g_{rr}&=&f^{-1} \left[1+\epsilon F_1 (1-3 \cos^2\theta)\right], \\
	k(r,\theta)&=&1-\epsilon F_2(1-3 \cos^2\theta).
	\end{eqnarray}
In the spacetime metric the terms being proportional to $\epsilon$ provide the part being responsible for the quasi-Schwarzschild effects. One can easily check that in the case of $\epsilon=0$ one recovers Schwarzschild spacetime. It is worth noting here that the condition $g^{rr}=0$ gives the location of an event horizon at $r_e=2 M$ being the same as in the case of the Schwarzschild black hole.

Using Wald method~\cite{Wald74} one can find the components of four vector potential of electromagnetic fields as
\begin{eqnarray}
A^\mu=\left(0, 0, 0, \frac{1}{2}B\right)\ .
\end{eqnarray}
Using the metric (\ref{metric}) one can write the covariant components as
\begin{eqnarray}
A_\mu=\left\{0, 0, 0, \frac{1}{2} B h(r,\theta) r^2 \sin^2\theta \right\}\ .
\end{eqnarray}
Now one can find the expression for the magnetic field around a quasi-Schwarzschild compact object. The four velocity of the proper observer is given by
\begin{eqnarray}
\mathcal{U}^\alpha=\left\{ \left(f \left[1+\frac{1}{2}\epsilon  F_1 (1+3 \cos2\theta)\right]\right)^{-\frac{1}{2}},0,0,0\right\}\ .
\end{eqnarray}
Then the orthonormal components of the magnetic field with respect to the chosen frame takes the following form~\cite{Narzilloev19}
\begin{eqnarray}\label{B}
B^{\hat{r}}&=&B \cos \theta   \frac{ 1+\epsilon  F_2 (3
    \cos 2 \theta -1)}{1+ \frac{\epsilon}{2} F_2 (3 \cos 2 \theta +1)},
\\\nonumber
B^{\hat{\theta}}&=&B \sin \theta  \sqrt{f}\frac{ \sqrt{1+\frac{\epsilon}{2}  F_1 (1+3 \cos 2 \theta)} }{\sqrt{1+\frac{\epsilon}{2}F_2 (3 \cos 2 \theta +1)}}\\ &\times& \frac{1+ \frac{\epsilon}{4}\left[rF_2'+2 F_2\right](3 \cos 2 \theta +1)}{\sqrt{1-\frac{\epsilon ^2}{4}F_1^2 (3 \cos 2 \theta +1)^2}}, \\
	B^{\hat{\phi}}&=&0\ , 
\end{eqnarray}
where the prime $'$ denotes derivative over the radial coordinates.
One can easily see that in pure Schwarzschild spacetime it takes 
\begin{eqnarray}
B^{\hat{r}}&=&B \cos \theta\ ,\\
B^{\hat{\theta}}&=& B \sqrt{f} \sin \theta \ ,
\end{eqnarray}
and in the Newtonian weak field regime $M/r \rightarrow0$ the components of the magnetic field become
\begin{eqnarray}
B^{\hat{r}}=B \cos \theta\ , \qquad
B^{\hat{\theta}}= B \sin \theta \ ,
\end{eqnarray}
consistent with the Newtonian limit.

\subsection{Circular motion of charged test particle around quasi-Schwarzschild compact object}

Now, we study the equation of motion of a charged particle around a quasi-Schwarzschild compact object briefly. It is more convenient to use Hamilton-Jacobi equation of motion for particles orbiting around central objects which is the case here. In the presence of an external electromagnetic field the equation reads

\begin{equation}\label{HamJam}
g^{\alpha\beta}\left(\frac{\partial S}{\partial x^\alpha}+eA_\alpha\right)\left(\frac{\partial S}{\partial x^\beta}+eA_\beta\right)=-m^2\ ,
\end{equation}
with $e$ and $m$ being the electric charge and mass of the test particle, respectively.

The equation of motion (\ref{HamJam}) is not separable when the system is non-integrable. In this case, Eq.~(\ref{HamJam}) should be replaced with a Hamiltonian formalism~\cite{Kolos15,Takahashi2009ApJ,Kopacek2010ApJ,Kopacek2014ApJ,LiDan2019EPJP,YiMiao2020PhyS}. However, when one investigates test particle motion on the equatorial plane, Eq.~(\ref{HamJam}) can be expressed in the following separable form
\begin{eqnarray}\label{S}
S &=&- E t+ L \phi + S_{r}+S_{\theta}\ ,
\end{eqnarray}
where $E$ and $L$ define the energy and angular momentum of a test particle per unit mass, respectively. Thus, the equation of motion of a test particle with unit mass reads as
\begin{eqnarray}\nonumber
&&\frac{\left(\mathcal{L}+\frac{e B}{2m}  r^2 (1-\epsilon  F_2)\right)^2}{r^2 (1-\epsilon  F_2)} +\frac{f}{1+\epsilon  F_1} \left(\frac{\partial S}{\partial r}\right)^2
\\ 
&& -\frac{\mathcal{E}^2}{f(1-\epsilon  F_1))} =-1\ ,
\end{eqnarray}
where ${\cal E}=E/m$ and ${\cal L}=L/m$ are specific energy and angular momentum of a test particle, respectively.

\begin{figure*}[t!]
	\begin{center}
		a.
		\includegraphics[width=0.45\linewidth]{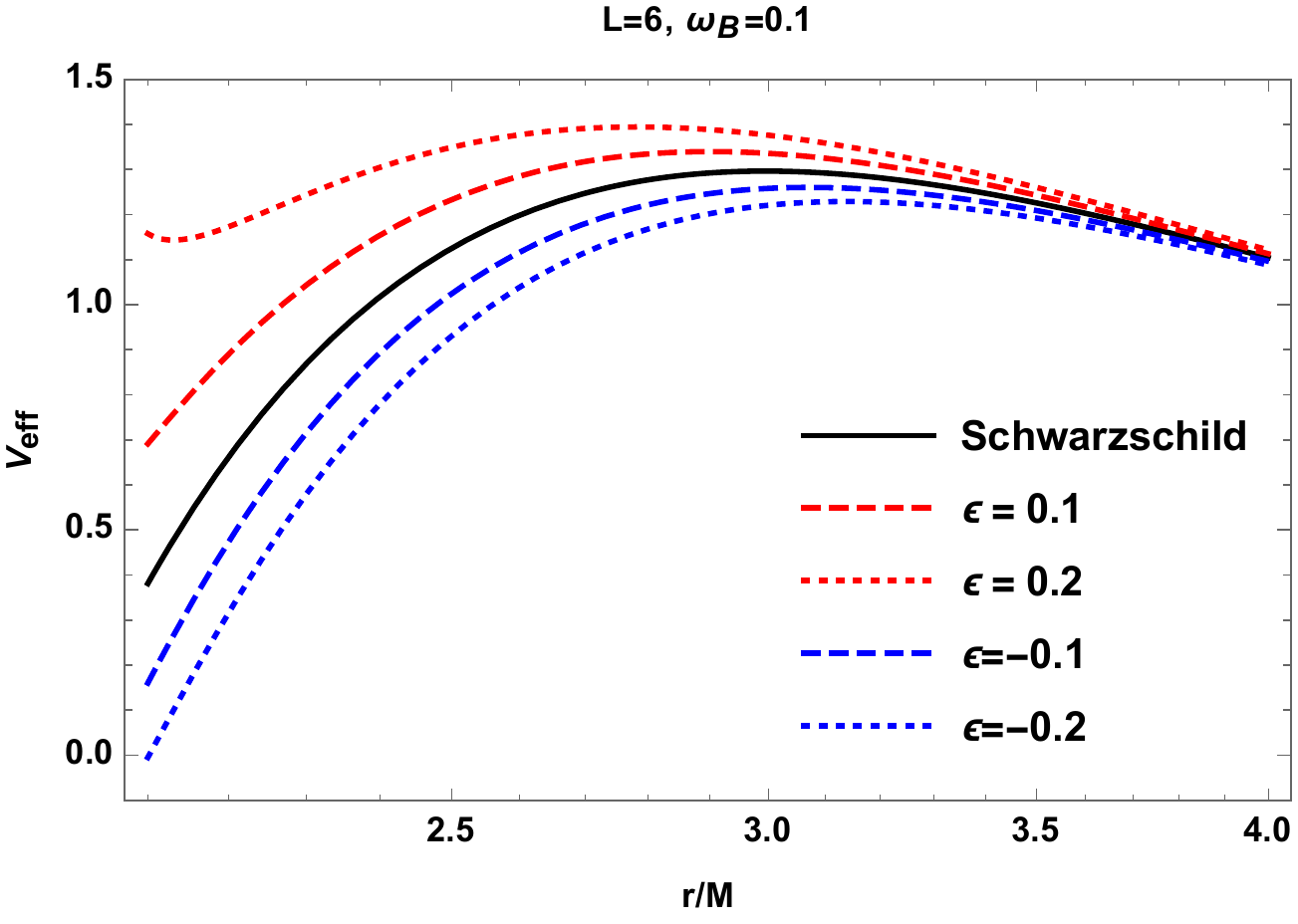}
		b.
		\includegraphics[width=0.45\linewidth]{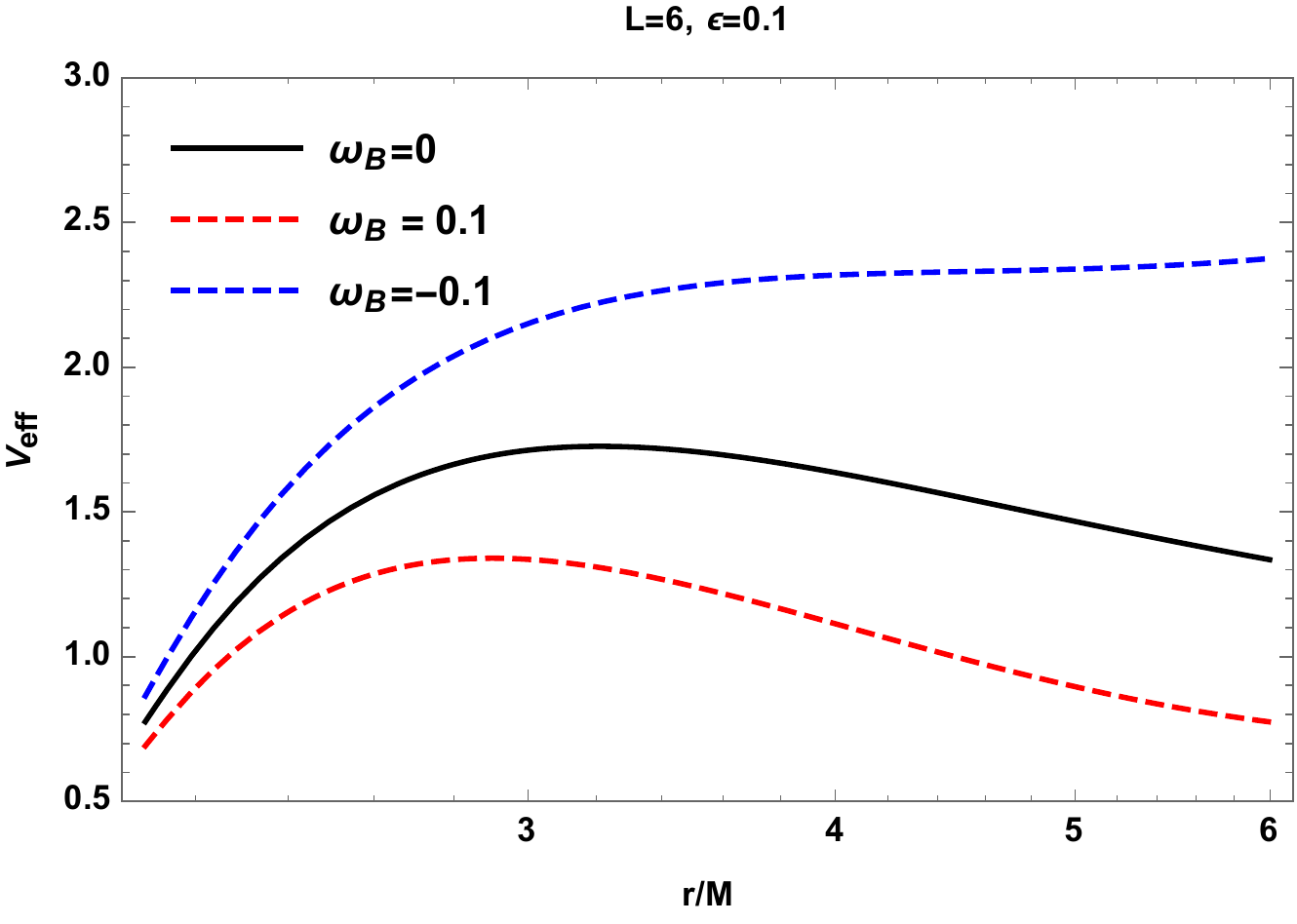}
	\end{center}
	\caption{Effective potential as a function of radial coordinate $r$ for the case $M=1$. Left panel corresponds for variation of deformation parameter. Right panel is for variation of magnetic interaction. \label{Veff}}
\end{figure*}

\textcolor{black}{For test particles moving at the equatorial plane ($\theta=\pi/2$) one can obtain the effective potential from the radial motion} %where $\dot{r}=0$ using the following standard relation
\begin{eqnarray}
\dot{r}^2=\mathcal{E}^2-V_{\rm eff}\ ,
\end{eqnarray}
that reads as
\begin{eqnarray}
V_{\rm eff}&=&f \left(1-F_1 \epsilon \right) \left\{1+\left(\frac{\mathcal{L}}{h(r,\frac{\pi}{2}) r^2}-\omega_B \right)^2\right\},
\end{eqnarray}
with $\omega_B=e B/(2mc)$ defining the cyclotron frequency of a charged particle which corresponds to the interaction between electrically charged particle and external magnetic field. The radial dependence of such effective potential is shown in Fig.~\ref{Veff}. It is clearly seen that the increase in deviation parameter also increases the effective potential while magnetic field has an opposite action as it was shown in Refs.\cite{Kolos15,Tursunov16,Stuchlik16}

\begin{figure*}[ht!]
	\begin{center}
		\includegraphics[width=0.45\linewidth]{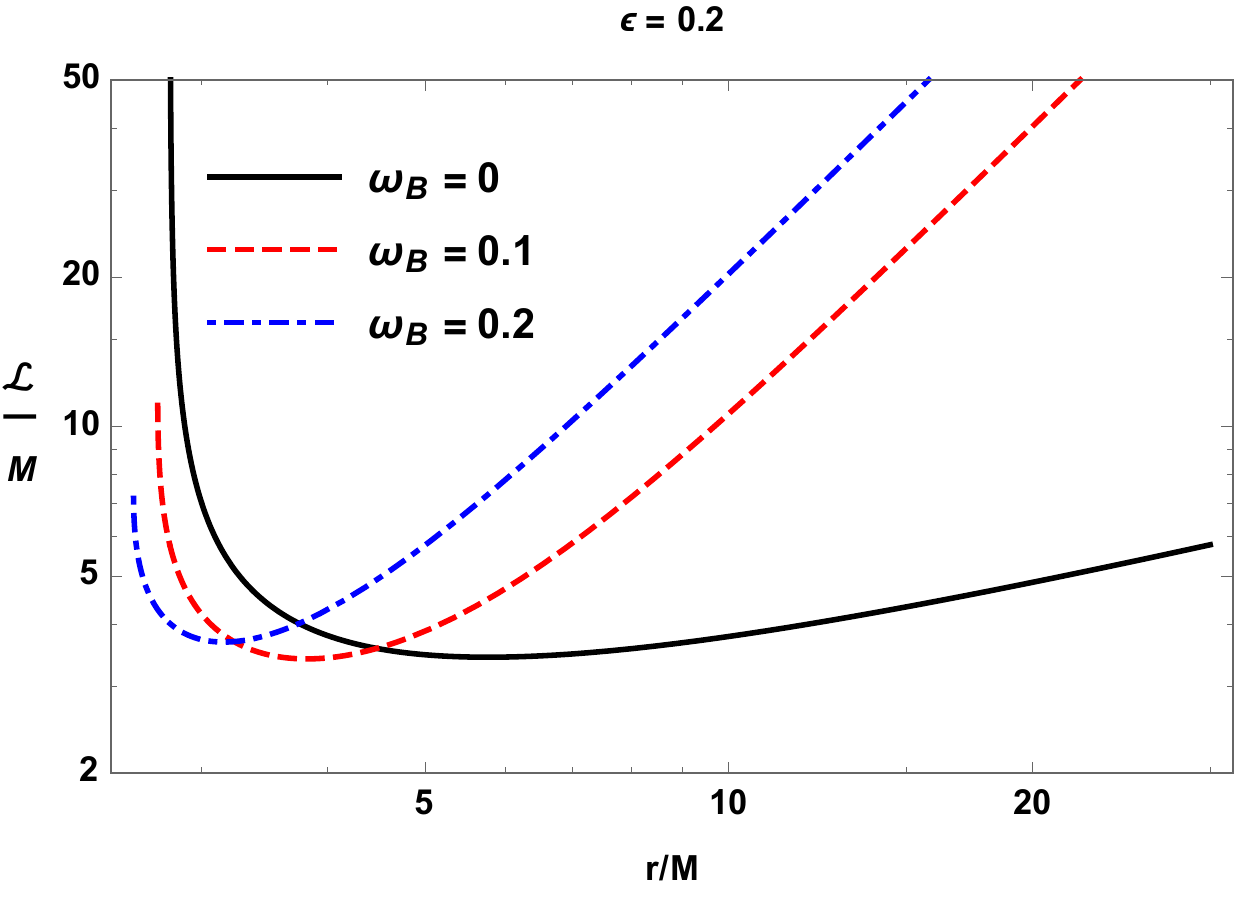}
		\includegraphics[width=0.45\linewidth]{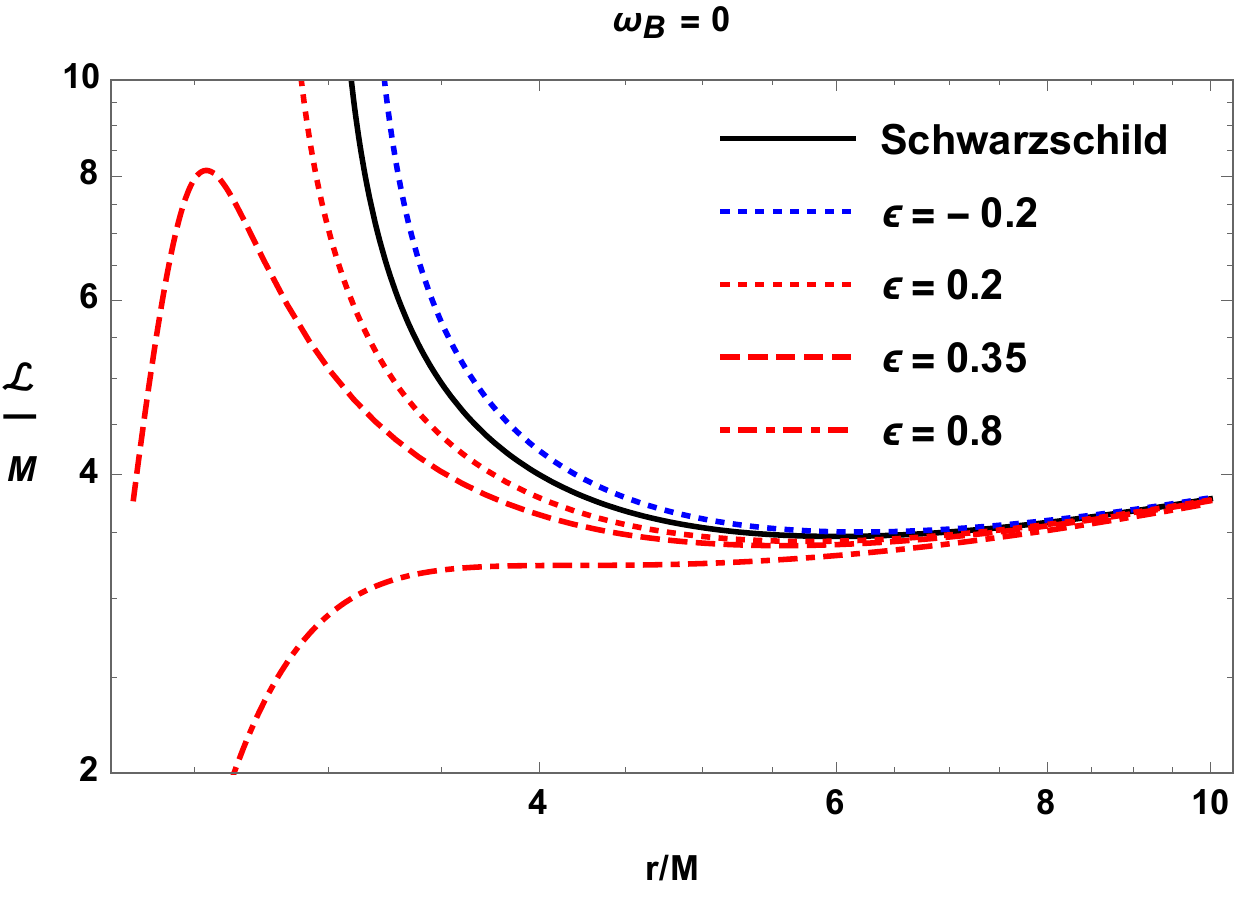}
	\end{center}
	\caption{Radial dependence of the angular momentum of a test particle for the different values of magnetic interaction (left panel) and deformation (right panel).\label{L}}
\end{figure*}

For a circular motion of a particle at equatorial plane one can set following standard conditions
\begin{eqnarray}\label{circle}
V_{\rm eff}(r)={\cal E}^2, \, V_{\rm eff}'(r)=0,  
\end{eqnarray}
which results the angular momentum of a charged test particle to have the radial dependence as presented in Fig. ~\ref{L}. We see the usual Schwarzschild shape for the line for which $\epsilon=0$ in the right panel. However, starting from some value of deviation parameter around $\epsilon\approx0.32$ it changes the shape of lines that causes them to have a maximum point at corresponding radius that vanishes starting from the case when $\epsilon\approx0.8$. We will come back to this point later in the next subsection where it contains an important description on defining ISCO radius.

\subsection{Innermost stable circular orbits. Quasi-Schwarzschild versus Kerr black hole}

\begin{figure*}[ht!]
	\begin{center}
		\includegraphics[width=0.45\linewidth]{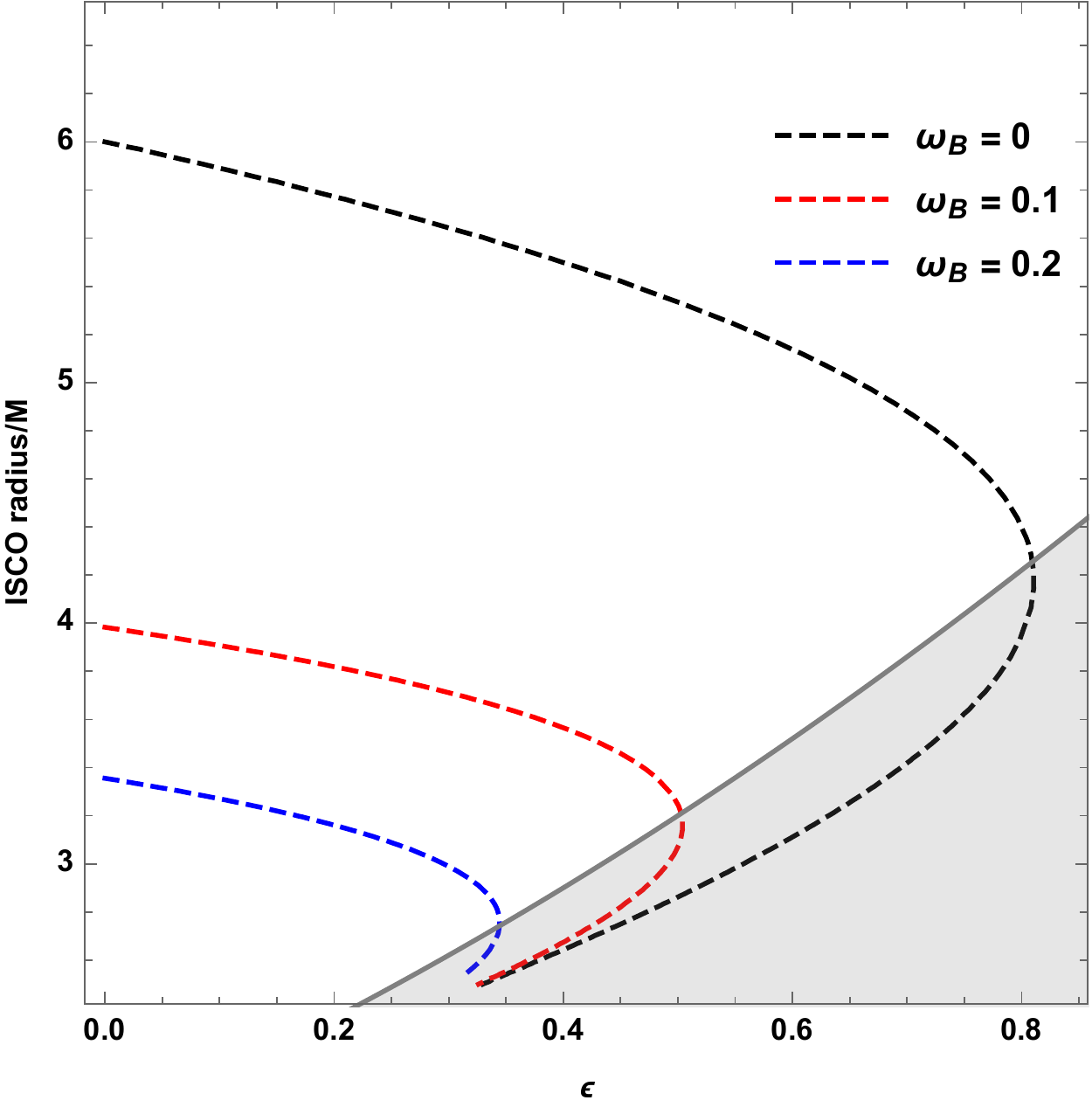}
		\includegraphics[width=0.465\linewidth]{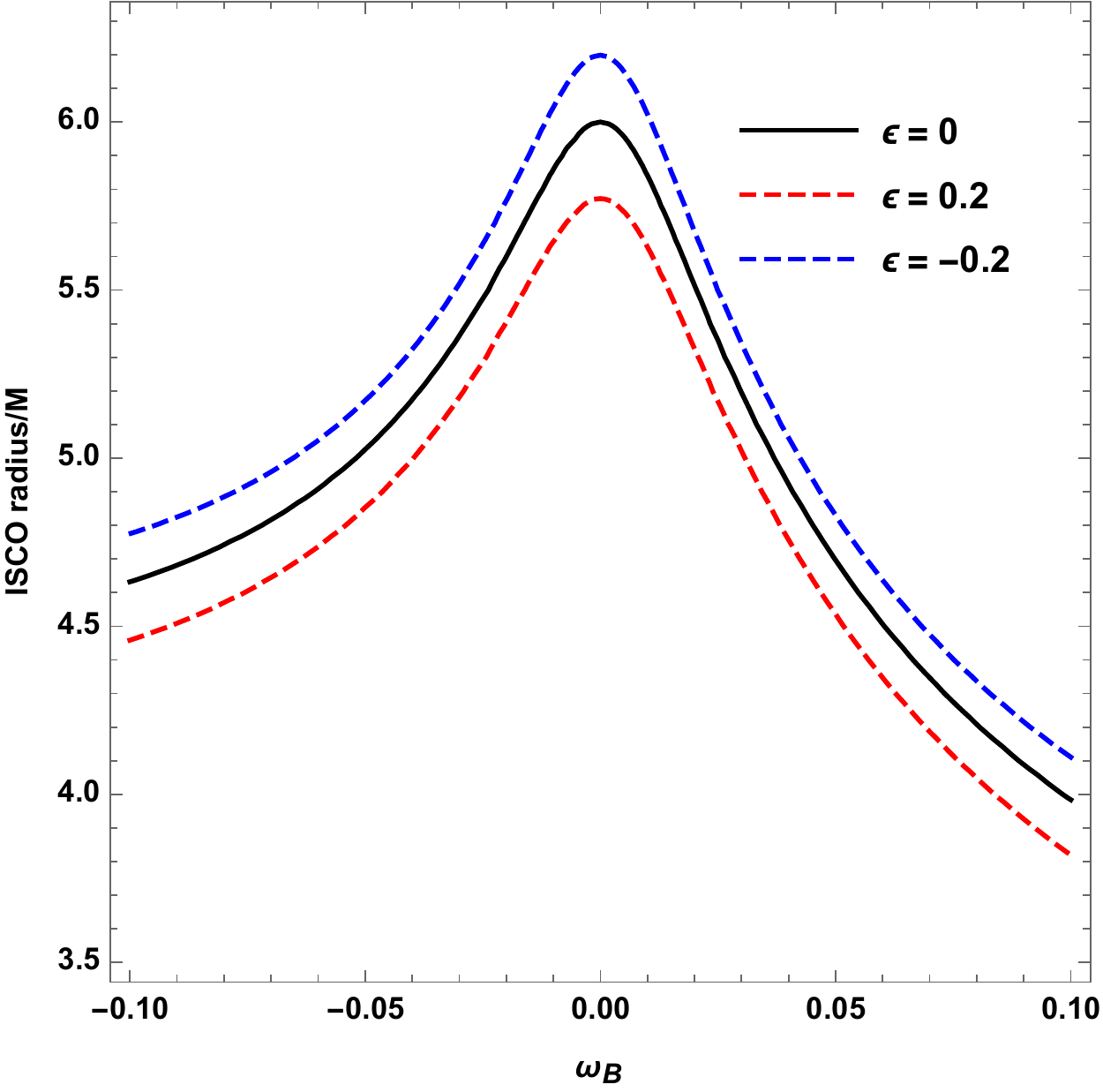}
	\end{center}
	\caption{ISCO radius of a test particle orbiting on equatorial plane of a quasi-Schwarzschild compact object for the different values of magnetic interaction (left panel) and deformation parameter (right panel).\label{isco}}
\end{figure*}

In this subsection we investigate so called innermost stable circular orbits (ISCO) around a quasi-Schwarzschild compact object immersed in an asymptotically uniform magnetic field. Based on the obtained results we will try to answer to the question, how the parameters $\epsilon$ and $B$ can mimic the rotation parameter $a$ of well known Kerr solution. The idea is, if the parameters mentioned can mimic the rotation parameter $a$ of a Kerr black hole, then for the same ISCO radius one can get a correspondence constraint between rotation parameter and parameters $\epsilon$ and $B$. First, we investigate the relation between ISCO radius and parameters of interest. To do so, one can add additional requirement to the condition (\ref{circle}) that reads
\begin{eqnarray}\label{iii}
V_{\rm eff}''(r)=0\ .
\end{eqnarray}
Taking into account these three conditions (together with (\ref{circle})) one might plot the dependence between ISCO radius and parameters $\epsilon$ and $B$ as plotted in Fig ~\ref{isco}. One can see from the graphs that the increase of the both parameters $\epsilon$ and $B$ reduces the ISCO radius of a charged test particle. From the left panel it is clearly seen that if one increases the deviation parameter $\epsilon$ up to some value then starts to reduce  the ISCO radius which becomes smaller and smaller instead of taking its initial values before reducing. To make the situation clear one should take into account the condition on angular momentum of a test particle that says that for particle to move on the last stable circular orbit its angular momentum should have a minimum on that orbit radius. Using this condition one can refer to the graph of dependence between angular momentum of a test particle on circular orbit radius as in the right panel of Fig. ~\ref{L}. We have mentioned this point in the previous subsection saying that the angular momentum for the absence of external magnetic field would have maximum points when the deviation parameter is between $0.32-0.8$. But, when we have plotted the Fig. ~\ref{isco} we just used the case when the angular momentum of a test particle has an extremum. So, we need to exclude such maximums from these extremum points which results that one needs to erase the lower part of ISCO lines starting from the turning points which is shown with shaded region in the left panel of Fig. ~\ref{isco}.
For the case of magnetic field the situation is typical as expected, i.e. if one increases the magnetic field the Lorentz force becomes stronger which makes the ISCO radius smaller.

\begin{figure*}[ht!]
	\begin{center}
		\includegraphics[width=0.48\linewidth]{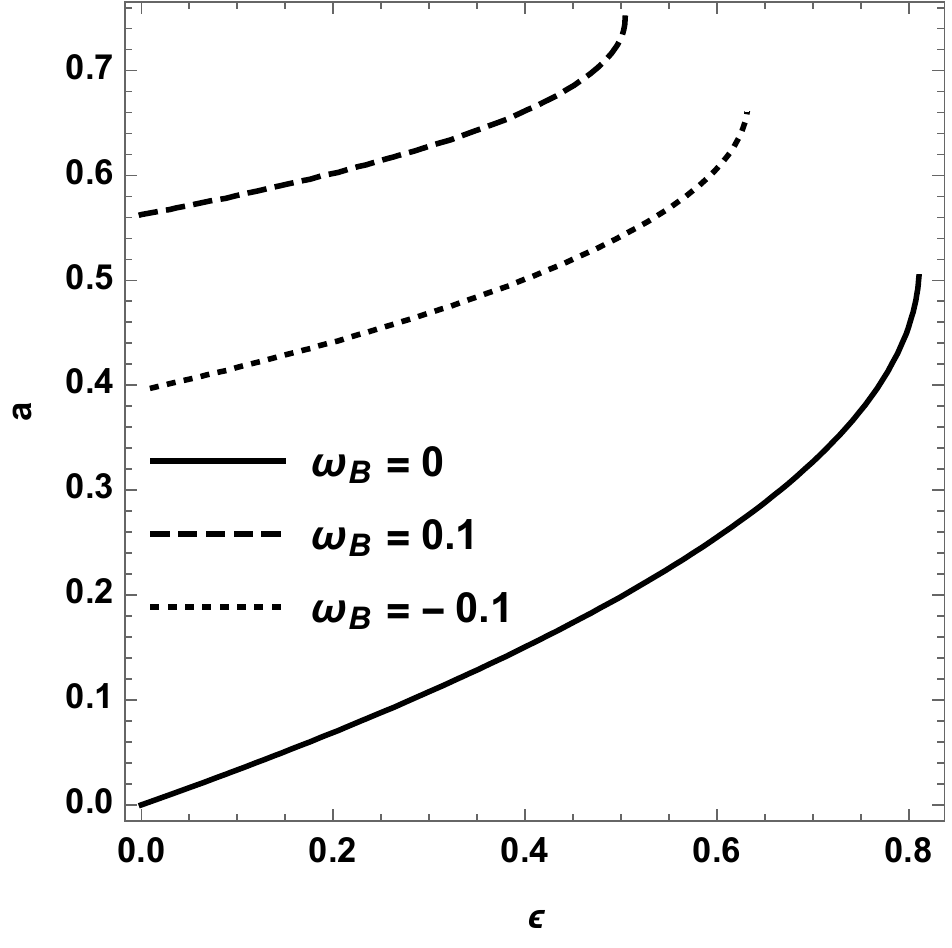}
		\includegraphics[width=0.48\linewidth]{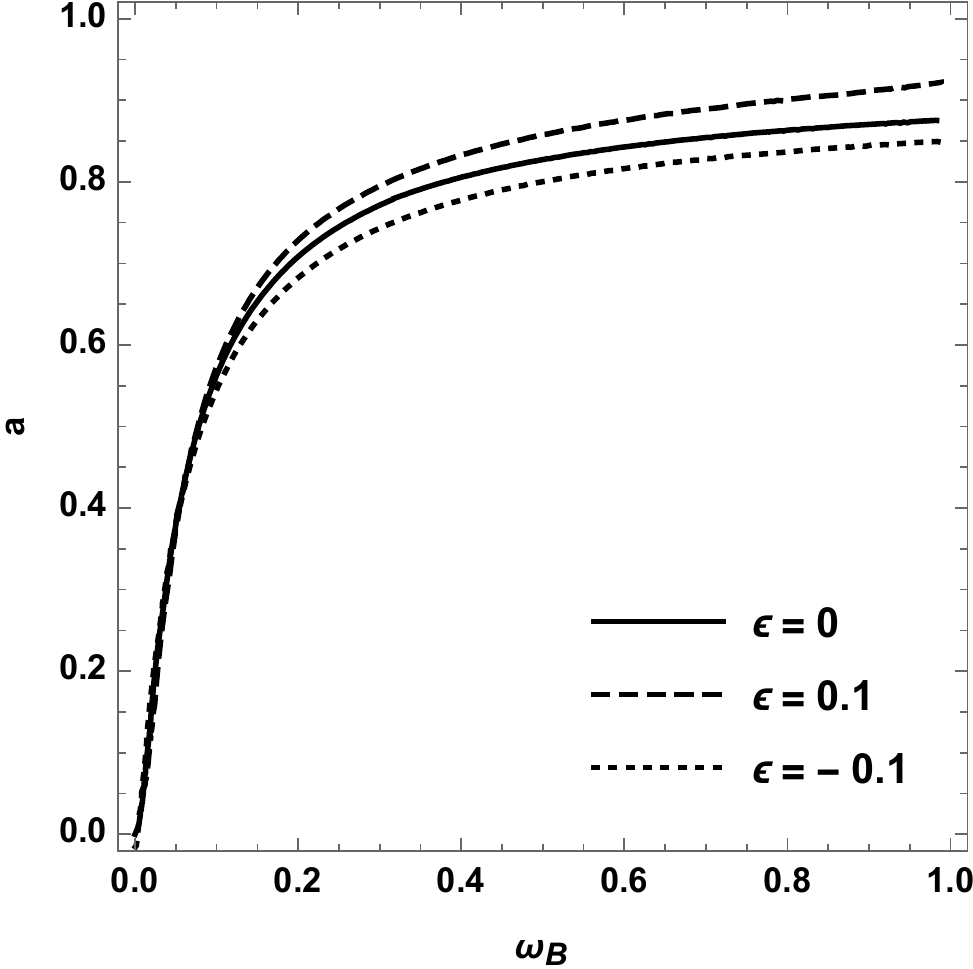}
	\end{center}
	\caption{Degeneracy plot that shows the correspondence between of dimensionless rotation parameter $a$ of a Kerr metric with $\epsilon$ and $B$ parameters. \label{deg}}
\end{figure*}

Finally, we plan to answer the question stated in the beginning of this subsection, how well the parameters $\epsilon$ and $B$ can mimic the rotation parameter $a$ of Kerr metric? Knowing how the ISCO in the case of Kerr metric behaves under the influence of a rotation parameter one can plot the degeneracy between these parameters as shown in Fig. ~\ref{deg}. It was expected from the dependence of ISCO on the parameter $\epsilon$ that this parameter can not completely mimic the rotation parameter as the ISCO radius did not tend to $M$ which is the case for extremal rotation $a\rightarrow1$  in Kerr metric. Now it is one more time clearly seen that this parameter can only mimic the rotation parameter up to approximately $\approx0.5$ when the external magnetic field is absent. One can however see that the magnetic parameter itself can mimic the rotation parameter up to $\approx0.88$. It is also seen from the right panel that in the presence of both, magnetic field and deviation parameter the mimic range exceeds $a>0.9$ being competitive with the rapidly rotating Kerr spacetime.

\begin{figure}[ht!]
	\begin{center}
		\includegraphics[width=0.90\linewidth]{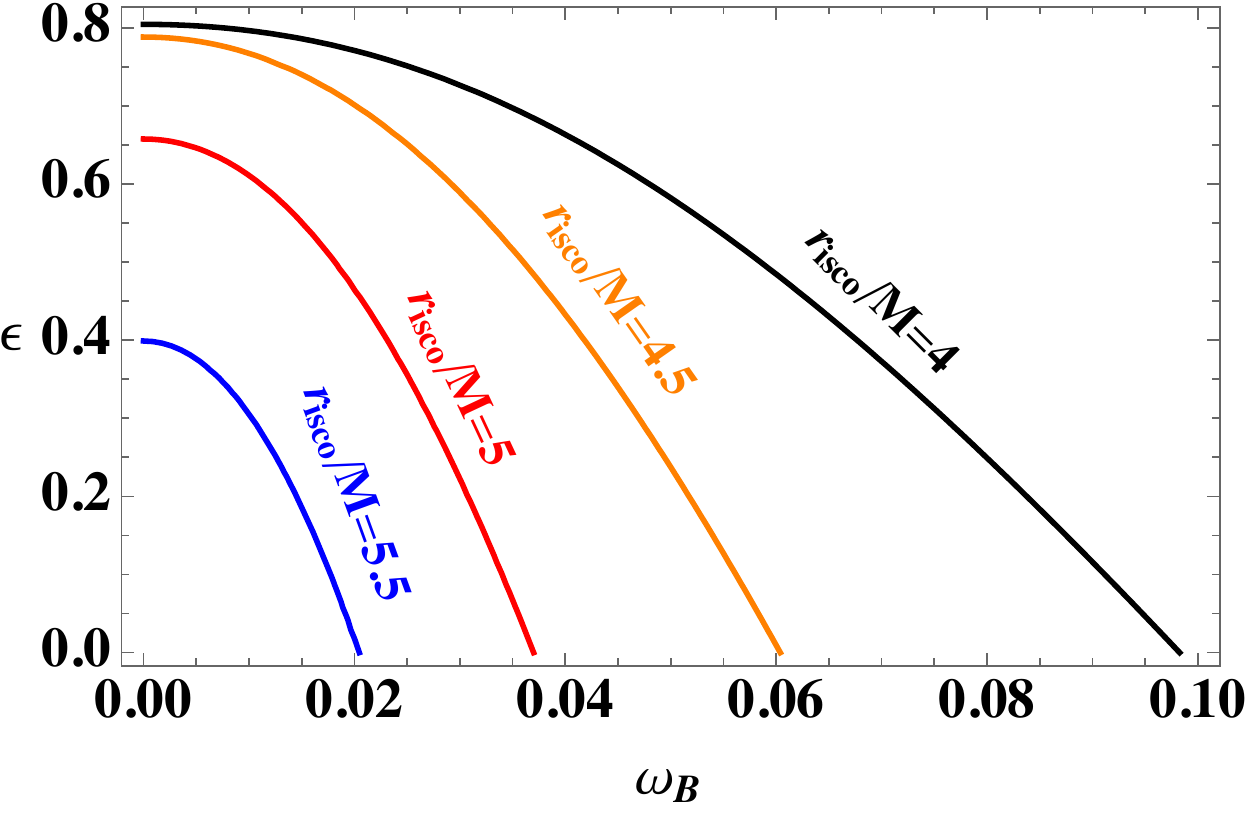}
	\end{center}
	\caption{Degeneracy plots for $\epsilon$ and $\omega_B$ for given values of ISCO radii.\label{e_B}}
\end{figure}

The degeneracy plot between deviation parameter $\epsilon$ and magnetic coupling parameter $\omega_B$ for a few fixed values of ISCO radius is illustrated in Fig. ~\ref{e_B}. We see that magnetic interaction has considerably stronger effect than the effect of deformation of a spacetime.

\section{Magnetic dipole motion. Quasi-Schwarzschild versus Kerr black hole \label{sectt3}}

In this section we focus on magnetic dipole motion around a quasi-Schwarzschild compact object immersed in an asymptotically  uniform magnetic field. The Hamilton-Jacobi equation of motion of magnetic dipole takes the following form

\begin{eqnarray}\label{H-J}
g^{\mu \nu}\frac{\partial {\cal S}}{\partial x^{\mu}} \frac{\partial {\cal S}}{\partial x^{\nu}}=-\Bigg(m-\frac{1}{2} D^{\mu \nu}F_{\mu \nu}\Bigg)^2\ .
\end{eqnarray}
Here, $D^{\mu\nu}$ is the antisymmetric polarization tensor which defines the electrodynamic properties of the particle. 
We assume that particle is electrically neutral $q=0$ and the polarization tensor is only described by magnetic moment $\mu$ itself. It is worth to note here that it is also possible to investigate in alternate way the motion of the magnetic dipole with an electric charge where one just needs to take the left hand side of Eq.\eqref{HamJam} instead of the one in Eq.\eqref{H-J}. However, in this work we are aimed to apply the motion of magnetic dipoles to magnetized neutron stars orbiting around supermassive black holes where neutron star can be treated as electrically neutral test particle with nonvanishing magnetic dipole moment. Since the mass of typical supermassive black hole is much greater than the mass of typical neutron star this allows us to take the neutron star as a test particle moving in the spacetime of the former one. Therefore hereafter we focus on the motion of electrically neutral magnetic dipole only.
In this case the components of this tensor can be written as~\cite{Narzilloev20c}
\begin{eqnarray}\label{D1}
D^{\mu\nu}=\eta^{\mu\nu\alpha\beta} u_{\alpha} \mu_{\beta} ,
\end{eqnarray}
that satisfies the following condition
\begin{eqnarray}\label{D2}
D^{\mu\nu} u_\nu=0 ,
\end{eqnarray}
where $\mu_\alpha$ describes the four magnetic momentum of a magnetic dipole.
$F_{\mu\nu}=\partial_\mu A_\nu-\partial_\nu A_\mu\ $ is the electromagnetic field tensor which can also be written in terms of the components of electromagnetic field as
\begin{eqnarray}\label{F}
F_{\mu\nu}=-\eta_{\mu\nu\alpha\beta} B^\alpha u^\beta+2 u_{[\mu} E_{\nu]}
\end{eqnarray}
where $\eta_{\alpha \beta \sigma \gamma}$ is the pseudo-tensorial form of the Levi-Civita symbol $\epsilon_{\alpha \beta \sigma \gamma}$ defined as
\begin{eqnarray}
\eta_{\alpha \beta \sigma \gamma}=\sqrt{-g}\epsilon_{\alpha \beta \sigma \gamma}\ , \qquad \eta^{\alpha \beta \sigma \gamma}=-\frac{1}{\sqrt{-g}}\epsilon^{\alpha \beta \sigma \gamma}\ ,
\end{eqnarray}
with $g={\rm det|g_{\mu \nu}|}=-r^4\sin^2\theta$ for spacetime metric (\ref{metric}) and 
\begin{eqnarray}
\epsilon_{\alpha \beta \sigma \gamma}=\begin{cases}
+1\ , \rm for\  even \ permutations\ ,
\\
-1\ , \rm for\  odd\  permutations\ ,
\\
\ \ 0 \ , \rm for\ the\ other\ combinations\ .
\end{cases}
\end{eqnarray}
Being contracted with (\ref{D1}) and using (\ref{D2}) with (\ref{F}) one can write
\begin{eqnarray}\label{DF}
D^{\mu\nu}F_{\mu\nu}= 2 \mu^\alpha B_\alpha=2\mu^{\hat{\alpha}} B_{\hat{\alpha}}\ .
\end{eqnarray}
For simplicity we consider the magnetic interaction between the magnetic dipole and external magnetic field to be weak enough (due to weakness of the external test magnetic field), so we can use the approximation  $\left({\cal D}^{\mu \nu}{\cal F}_{\mu \nu} \right)^2  \to 0$.
It is expected that in a given external magnetic field the magnetic momentum of a particle aligns along this external field. If one assumes a particle moving at equatorial plane ($\theta=\pi/2$) then since this magnetic field has only normal component $B^{\hat{\theta}}$ to this equatorial plane and so does the magnetic moment $\mu^{\hat{\theta}}$ which is consistent with the lowest energy condition of the magnetic dipole. The scalar product (\ref{DF}) then becomes

\begin{eqnarray}
D^{\mu\nu}F_{\mu\nu}= 2\mu B \mathcal{A}
\end{eqnarray}
here $\mathcal{A}(r)$ defines the proportionality function that taking into account expression (\ref{B}) and for $B^{\hat{\theta}}$ reads

\begin{eqnarray}
\mathcal{A}=\mathcal{A}(r)=\sqrt{f\frac{1-\epsilon  F_1}{1-\epsilon F_2 }} \, \frac{1- \frac{\epsilon}{2}\left[rF_2'+2 F_2\right]}{\sqrt{1-\epsilon ^2F_1^2}}\ .
\end{eqnarray}

\begin{figure*}[ht!]
	\begin{center}
			\includegraphics[width=0.494\linewidth]{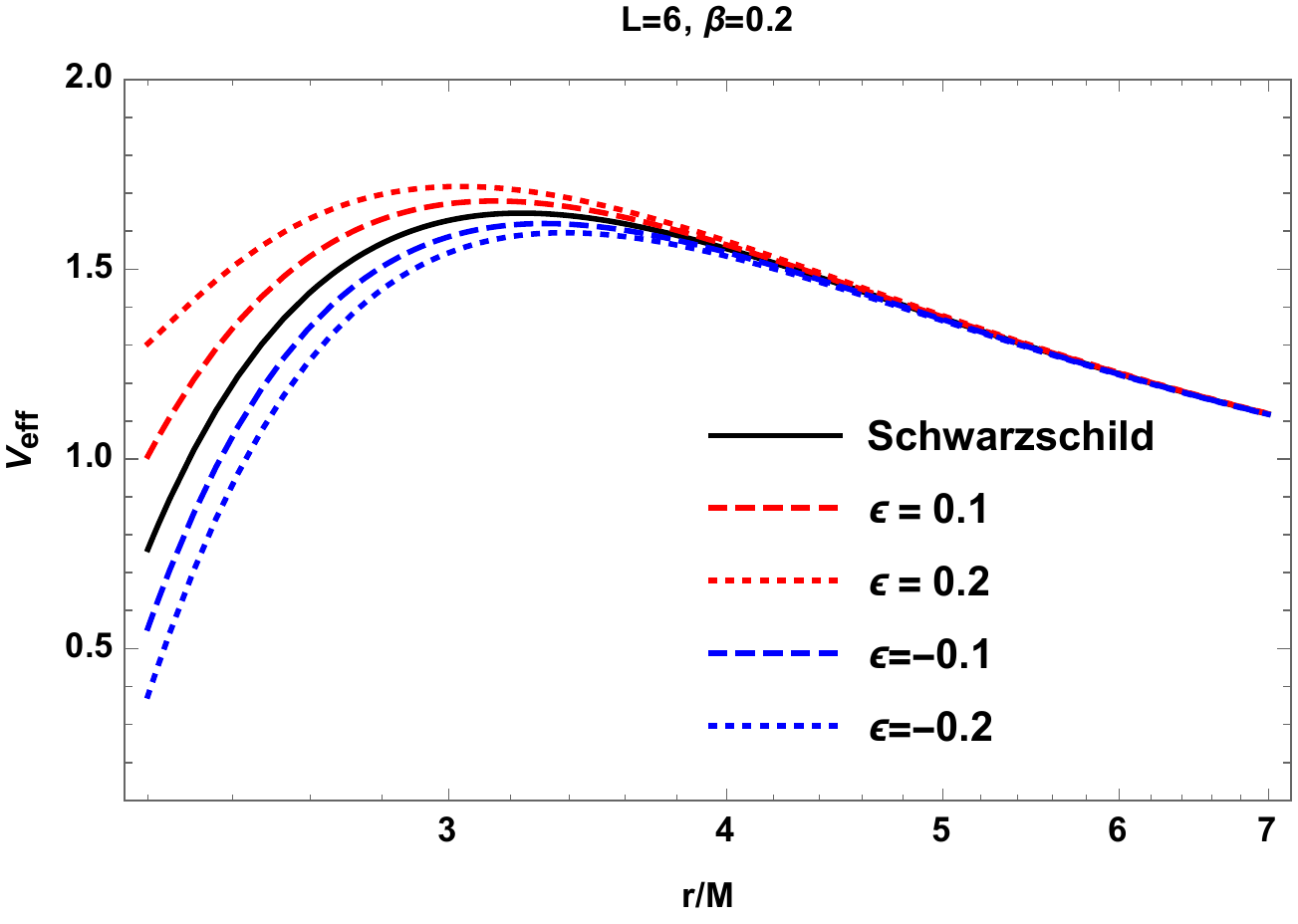}
		\includegraphics[width=0.494\linewidth]{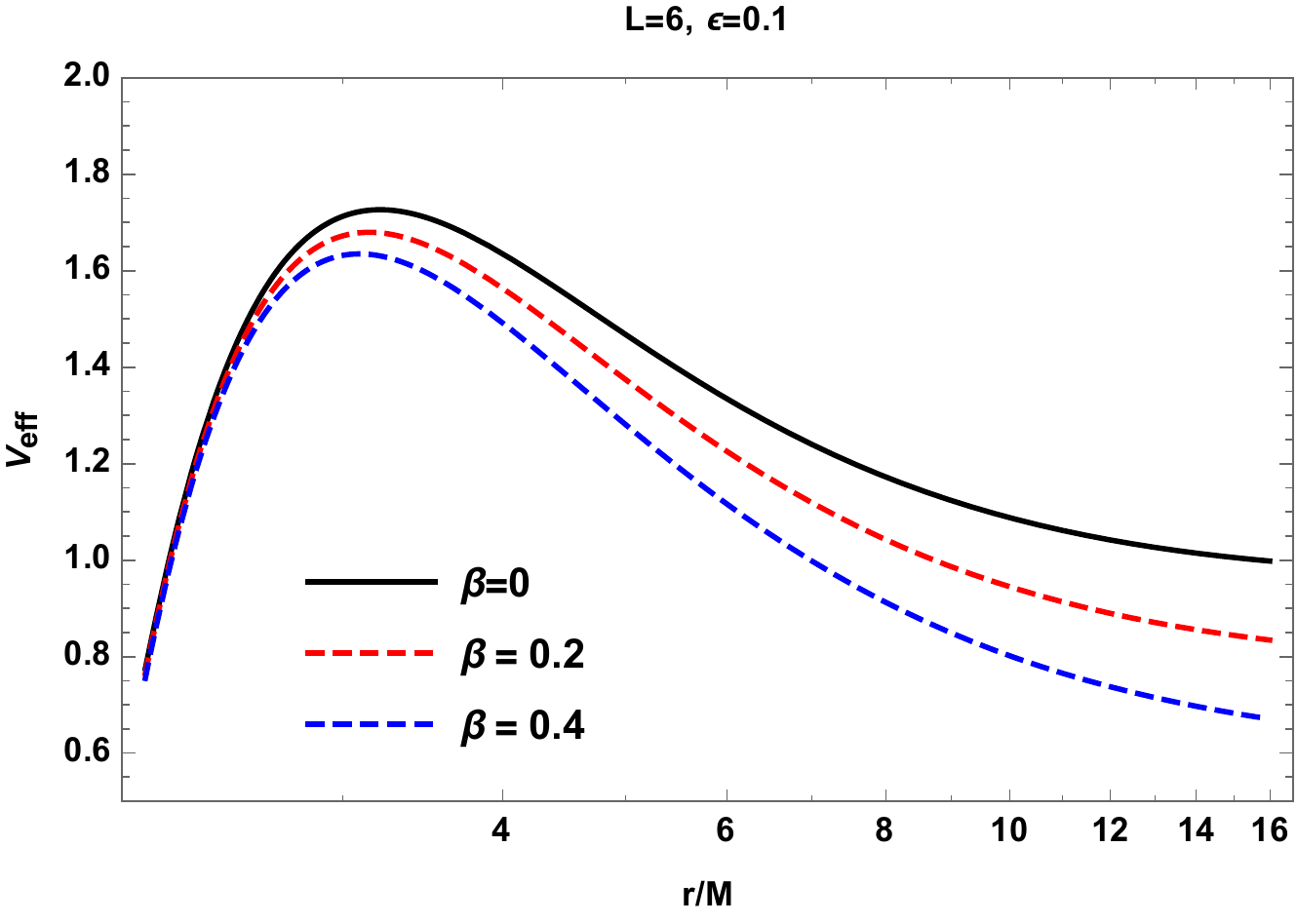}
	\end{center}
	\caption{Effective potential as a function of $r/M$ for different values of deviation (left panel) and magnetic interaction (right panel) parameters. \label{mVeff}}
\end{figure*}

Plugging the scalar product of $D^{\mu\nu}$ and $F^{\mu\nu}$ into the equation of motion (\ref{H-J}) one can find the effective potential at the equatorial plane as
\begin{eqnarray}
\nonumber
V_{\rm eff} &= & \left(1-\frac{2 M}{r}\right) \Big[1-\epsilon  F_1\Big]\\
&\times & \left\{\left(1-\beta 
   \mathcal{A}\right)^2+\frac{\mathcal{L}^2}{r^2 \left[1-\epsilon  F_2\right]}\right\}\ ,
\end{eqnarray}
here $\beta=2 \mu B/m$ is called magnetic coupling parameter that defines electromagnetic interaction between magnetic dipole and external magnetic field. In real astrophysical scenarios, for example in the case of typical neutron star orbiting around super-massive black hole with magnetic dipole moment $\mu=(1/2)B_{\rm NS}R^3_{\rm NS}$ the coupling magnetic parameter is 
\begin{eqnarray}\label{betaNS}
\beta \simeq \frac{11}{250}\left(\frac{B_{\rm NS }}{10^{12} \rm G}\right)\left(\frac{R_{\rm NS}}{10^6 \rm cm}\right)^3\left(\frac{B_{\rm ext}}{10\rm G}\right)\left(\frac{m_{\rm NS}}{M_{\odot}}\right)^{-1}, 
\end{eqnarray}
where $B_{\rm NS}$ is the magnetic field at the surface of neutron star,
$R_{\rm NS}$ and $m_{\rm NS}$ are radius and mass of neutron star, respectively. 
Radial dependence of the effective potential is plotted in Fig. ~\ref{mVeff}. We see that effective potential for magnetic dipole behaves similarly to the charged particle one.

From the same conditions (\ref{circle}) for the trajectory of the particle to be circular one can easily find the expressions for the angular momentum and the energy of the test particle that read
\begin{widetext}
\begin{eqnarray}
\mathcal{L}^2&=&\frac{r^3 (1-{\cal A} \beta ) \left(1- \epsilon F_2 \right)^2 \left\{(1-{\cal A} \beta ) \left[f' \left(1- \epsilon F_1 \right)- \epsilon f F_1'\right]-2 \beta  f {\cal A}' \left(1- \epsilon F_1\right)\right\}}{\left(1- \epsilon F_2 \right) \left[\left(1- \epsilon F_1 \right) \left(2 f-r f'\right)+f r \epsilon  F_1'\right]-f r \epsilon  \left(1- \epsilon F_1  \right) F_2'}\ , \\
\mathcal{E}^2&=&\frac{(1-\beta \mathcal{A}) f^2 \left(1- \epsilon F_1 \right)^2 \left\{(1-\beta \mathcal{A}) r \epsilon  F_2'+2 \left(1- \epsilon F_2 \right) \left[1-\beta  \left(\mathcal{A}+r\mathcal{A}'\right)\right]\right\}}{\left(1- \epsilon F_2  \right) \left[\left(1- \epsilon F_1 \right) \left(2 f-r f'\right)+ \epsilon f r   F_1'\right]-f r \epsilon  \left(1- \epsilon F_1 \right) F_2'} \ .
\end{eqnarray}
\end{widetext}

\begin{figure*}[ht!]
	\begin{center}
			\includegraphics[width=0.494\linewidth]{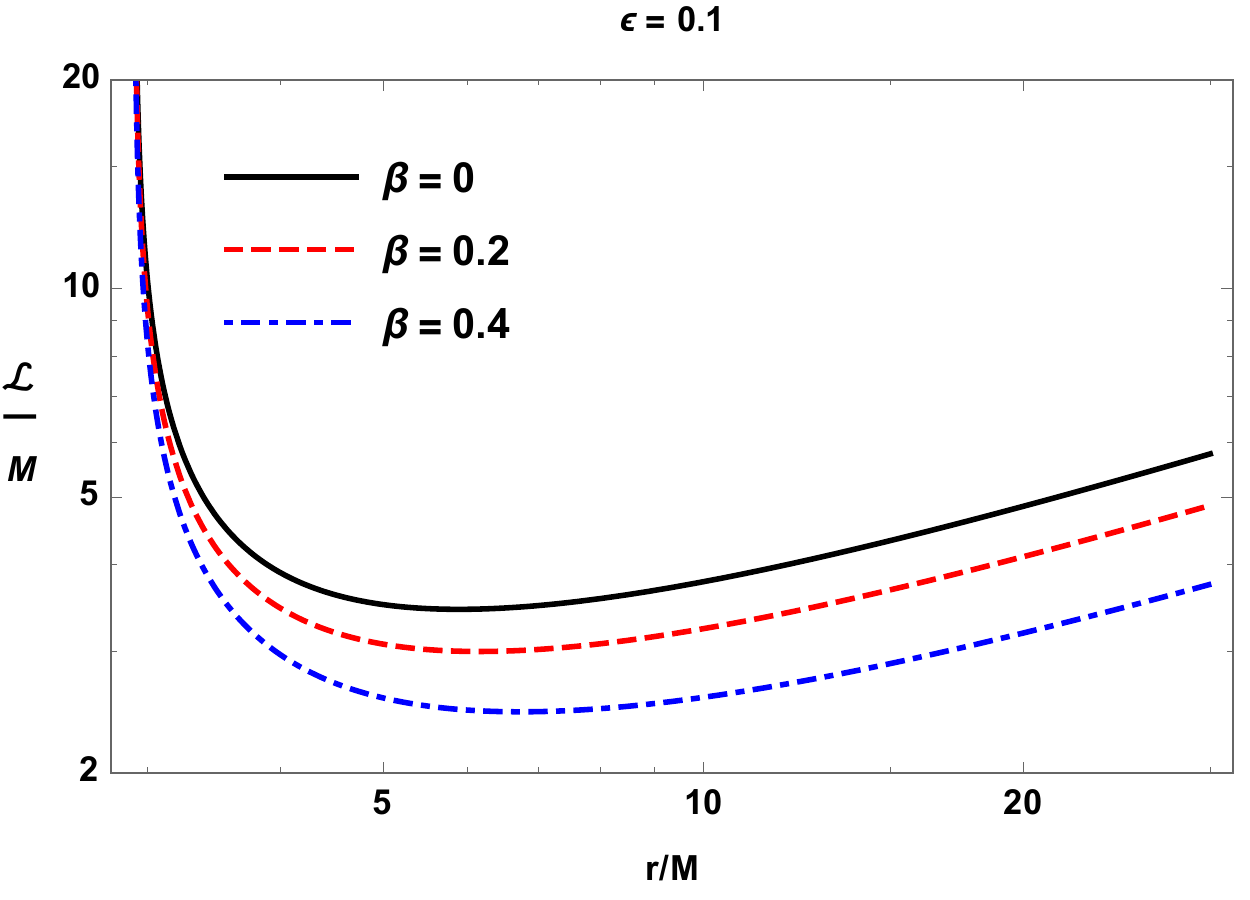}
			\includegraphics[width=0.494\linewidth]{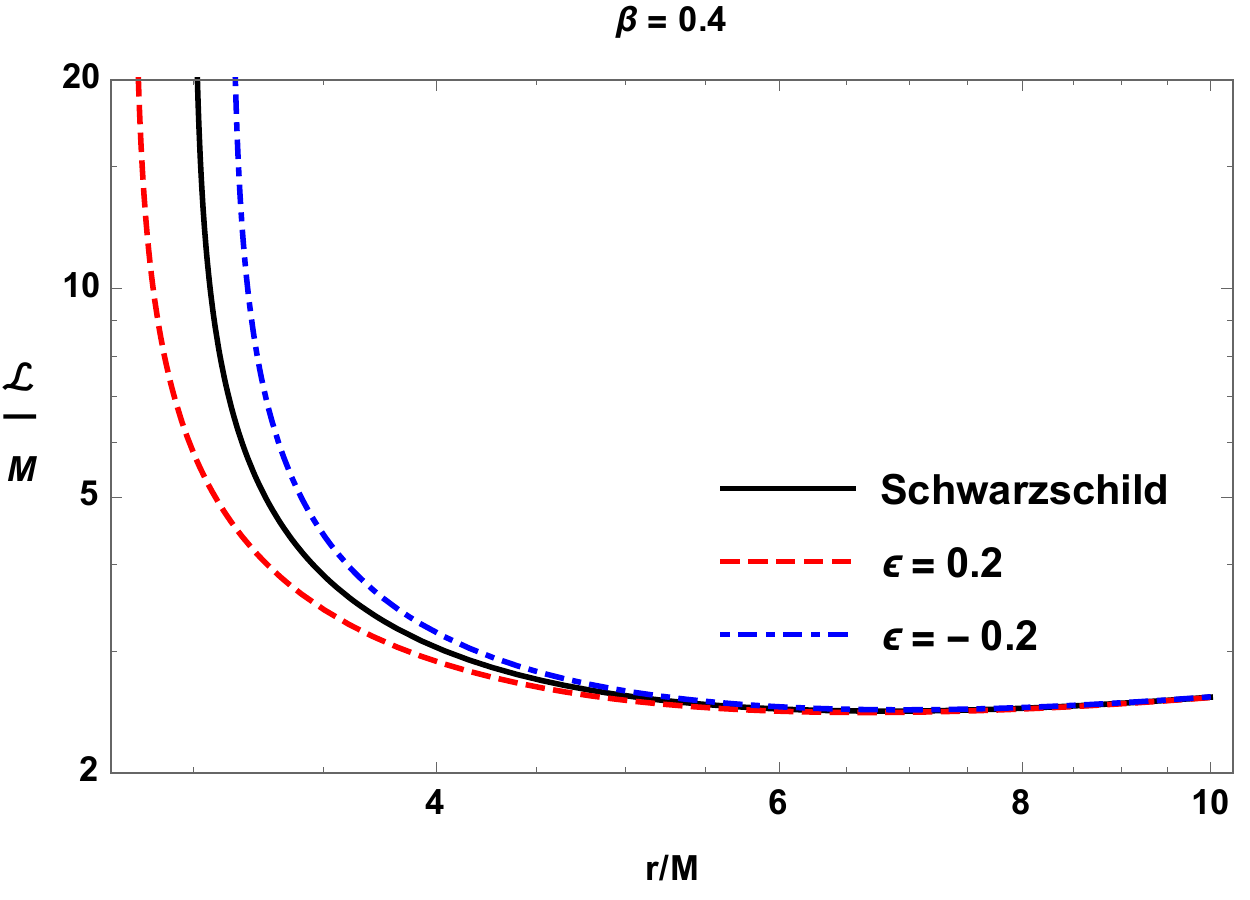}
	\end{center}
	\caption{Radial dependence of the angular momentum of a magnetic dipole for the different values of magnetic parameter $\beta$ (left panel) and deviation $\epsilon$ (right panel).\label{mL}}
\end{figure*}

The orbit of a particle moving at equatorial plane being circular makes the angular momentum of a particle to have the radial dependence as plotted in Fig. ~\ref{mL}. From the shift of the minimum of the lines one can state how the ISCO radius changes for the different values of magnetic coupling parameter and also for various values of deviation parameter. From the upper panel it comes out that if one increases the magnetic interaction between magnetic dipole and external magnetic field then it increases the ISCO radius as well.

\begin{figure*}[ht!]
	\begin{center}
		\includegraphics[width=0.494\linewidth]{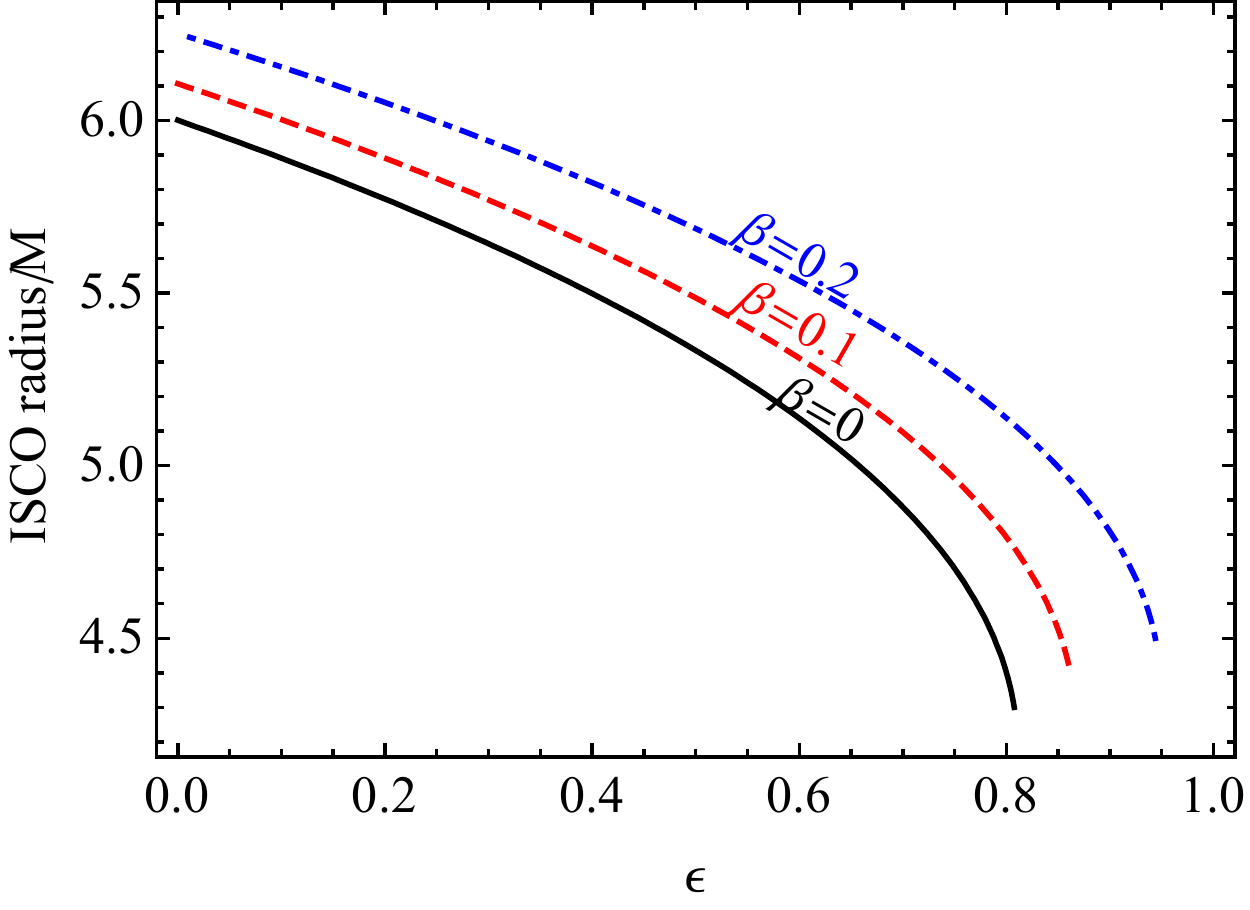}
		\includegraphics[width=0.494\linewidth]{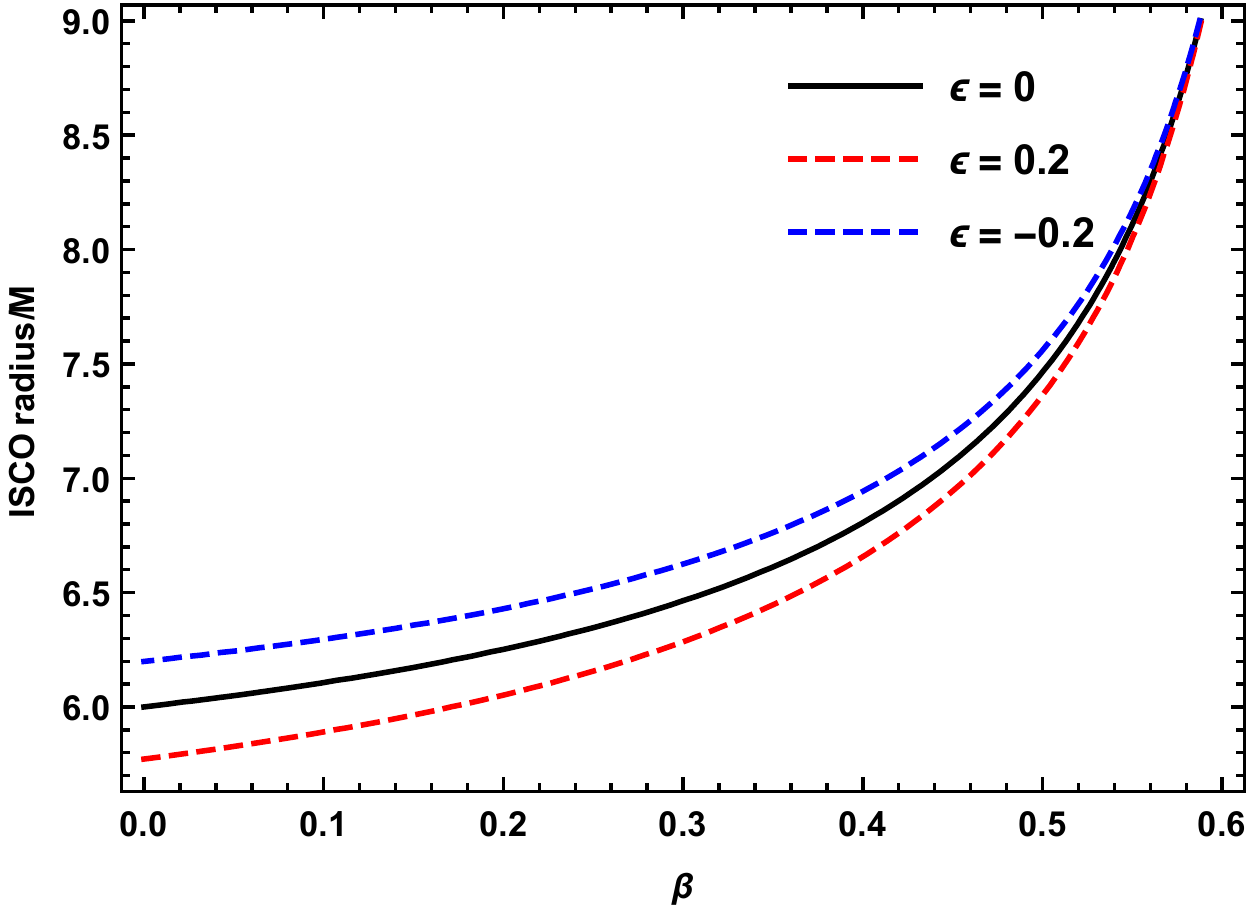}
	\end{center}
	\caption{Dependence ISCO radius of test particles orbiting at the equatorial plane around quasi-Schwarzschild black hole from the deviation for given $\beta$ (left panel) and magnetic coupling parameters for given $\epsilon$ (right panel).\label{misco}}
\end{figure*}

Having obtained the effective potential and the radial dependence for angular momentum one can now investigate ISCO for magnetic dipole moving at the equatorial plane. We can use either the condition given in Eq. (\ref{iii}) or for the angular momentum to have a minimum at ISCO radius. Then the dependence of ISCO radius from parameters $\epsilon$ and magnetic coupling parameter $\beta$ becomes as presented in Fig. ~\ref{misco}.

Figure \ref{misco} demonstrates the dependence of ISCO radius of magnetic dipole around Quasi-Schwarzschild black hole from the deviation (on the top panel) and magnetic coupling (on the bottom panel) parameters. We see from the plots that increasing the magnetic coupling parameter increases the ISCO radius. Moreover, it appears that it has a value around $\beta=2/3$ at which ISCO radius tends to infinity saying that no stable circular orbits can occur no matter how far the magnetic dipole is orbiting. In the upper panel we have cut the lower part of ISCO radius dependence due to the same reason as in the previous section which says that for chosen value of deviation parameter the angular momentum of a test particle can have both minimum and maximum where we should take only minimum points that are physically relevant.

\section{Astrophysical applications of the study\label{sectt4}}

One of the important and actual key issue in relativistic astrophysics is testing theory of gravity by study of test magnetic dipoles motion, in particular, in exploration of motion of neutron stars (pulsars and/or magnetars) treated as test magnetic dipoles around supermassive black holes (hereafter SMBH), that may give a possibility to test both gravitational and electromagnetic fields around a SMBH due to their accurate pulses in observations which may help to measure the distance through Doppler effect. In observations of such models it can be realized when neutron star could be found near the center of a galactic center. However, by now, it is quite difficult to find radio pulsars, due to Compton scattering of radio pulses in dense charged electron gas around the Sgr A*. The first and by now a single neutron star--magnetar called SGR 1745-2900 around Sgr A* has been discovered in 2013~\cite{Mori2013ApJ}. In our calculations we use the parameters of the magnetar treating it as a magnetic dipole orbiting Sgr A*. On other hand, theoretical problematic issue on the analysis of observational properties such as QPO, ISCO radius around the SMBH when the parameters of different alternate gravity reflect the similar effects on the properties in that cases it is impossible which gravity's effect plays dominant role. In fact that mostly astrophysical black holes are accepted as rotating black holes. Here we aimed to analyze ISCO radius comparing with the effects of spacetime deformation and spin of Kerr black hole when both provide the same value for ISCO radius of magnetic dipoles. Note that for comparison we consider quasi-Schwarzschild compact object immersed in an external asymptotically uniform magnetic field and Kerr black hole without magnetic field. We assume real astrophysical case of the magnetar orbiting the SMBH Sgr A*. Note also that one can consider a magnetic dipole as a neutral one in the absence of external magnetic field.           

The value of the magnetic coupling parameter $\beta$ for the magnetar SGR (PSR) J1745--2900 with the magnetic dipole moment $\mu \simeq 1.6\times 10^{32} \rm G\cdot cm^3$  orbiting the supermassive BH Sgr A* is~\cite{Mori2013ApJ}
 \begin{eqnarray}
 \beta_{\rm PSR\, J1745-2900}\simeq 0.716 \left(\frac{B_{\rm ext}}{10\rm G}\right)\ .
 \end{eqnarray}

ISCO radius of test particles around a rotating Kerr black hole is defined by the following expression 
for retrograde (+) and prograde (-) orbits~\cite{Bardeen72}
\begin{eqnarray}
r_{\rm isco}= 3 + Z_2 \pm \sqrt{(3- Z_1)(3+ Z_1 +2 Z_2 )} \ ,
\end{eqnarray}
where
\begin{equation} \nonumber
Z_1-1=\left(\sqrt[3]{1+a}+\sqrt[3]{1-a}\right)\sqrt[3]{1-a^2}\ ,
\quad
Z_2^2-Z_1^2=3a^2 \ .
\end{equation}

Now, in order to compare effects of spin and deviation parameters on ISCO radius we will provide the dependence of ISCO radius from deviation parameter $\epsilon$ and spin of Kerr black hole for the magnetic dipoles with the negative and positive values of the magnetic coupling parameters as $\beta=\pm 0.25$ and neutral particles, noting that when the direction of the magnetic dipole moment of the test particle aligns along the magnetic field the magnetic coupling parameter is positive, otherwise it is negative. 

\begin{figure*}[ht!]
	\begin{center}
		\includegraphics[width=0.45\linewidth]{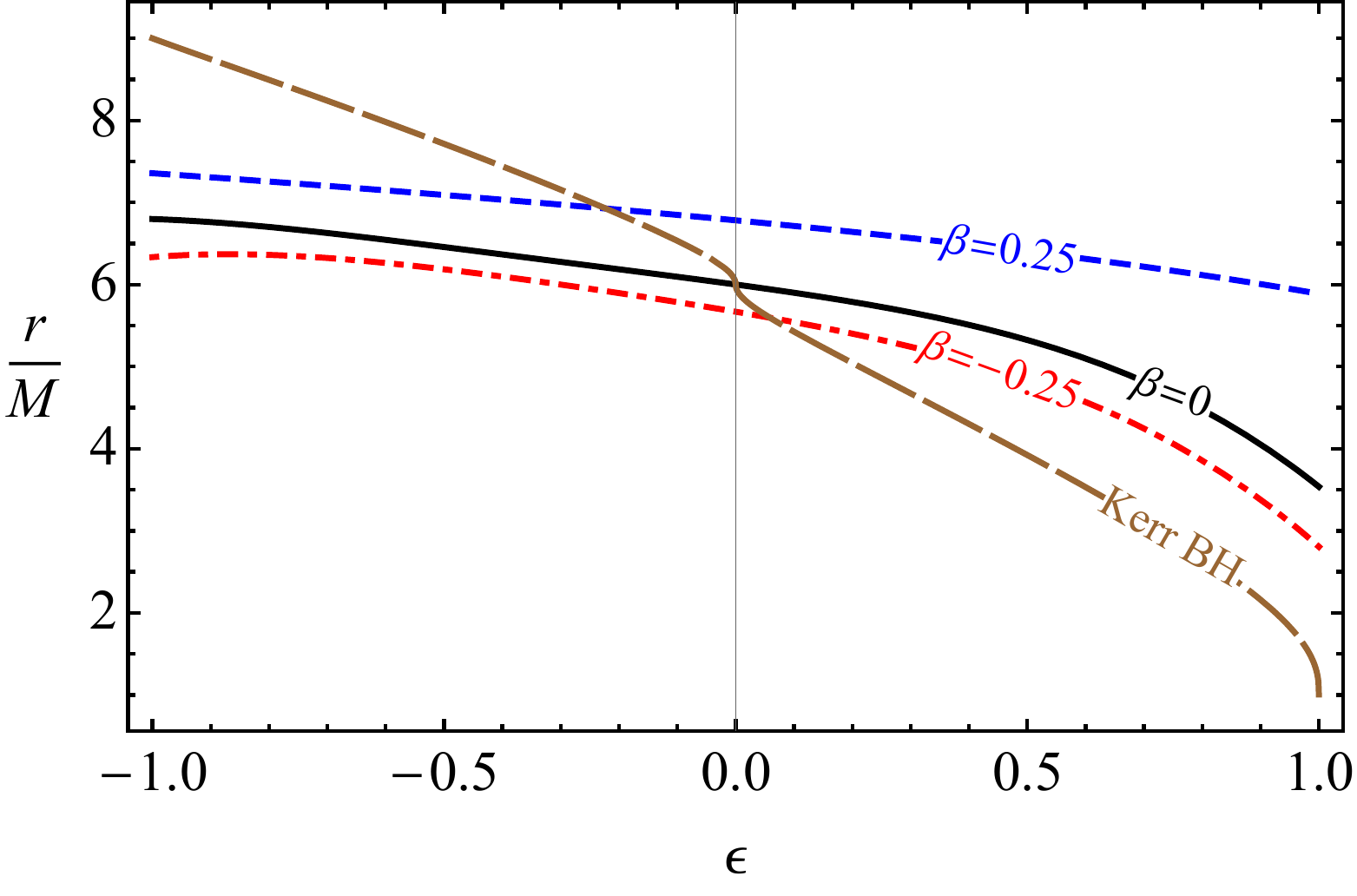}
		\includegraphics[width=0.47\linewidth]{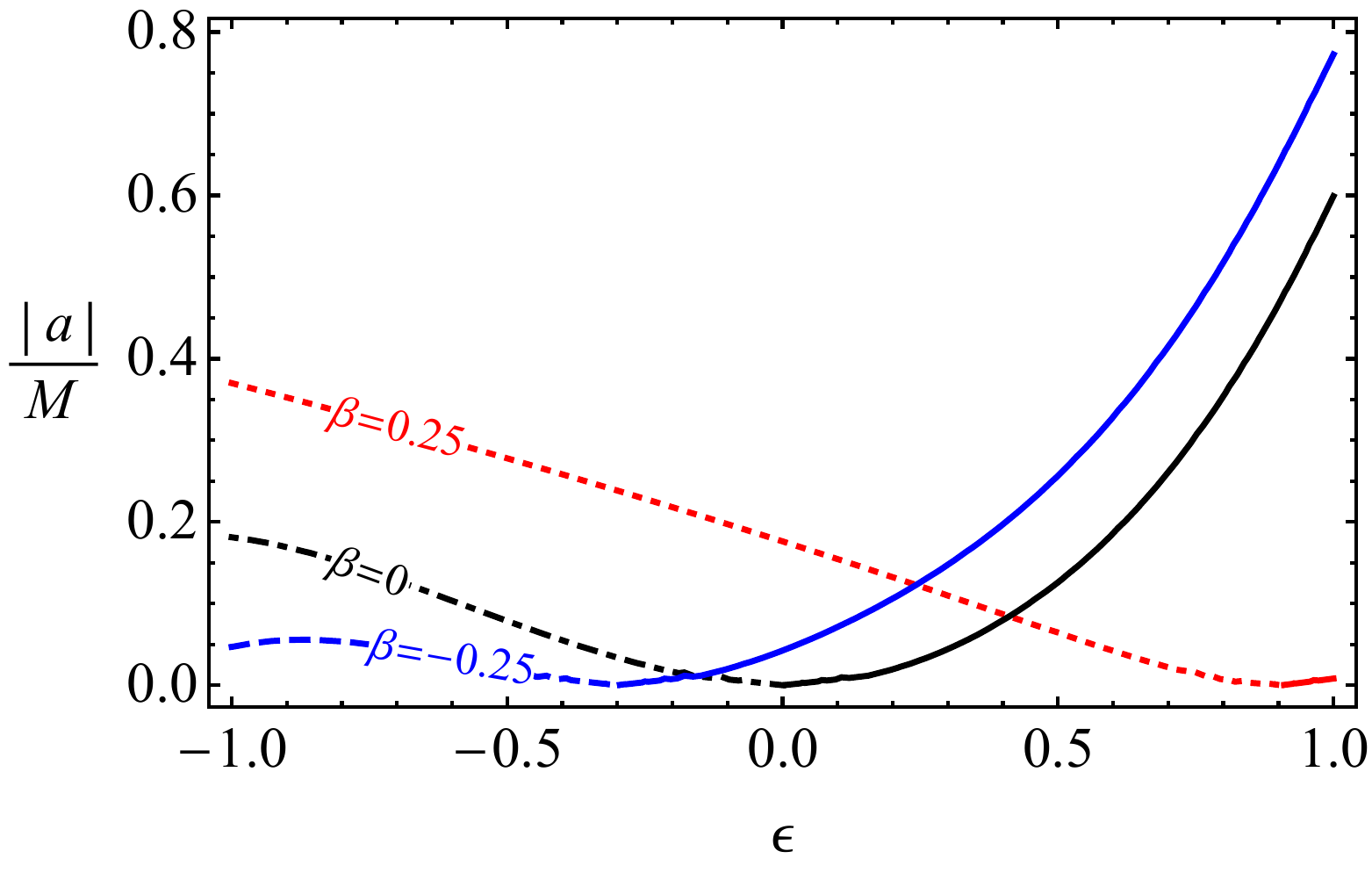}
	\end{center}
	\caption{Dependence of ISCO radius of magnetic dipoles from deviation parameter of quasi-Shwarzschild black hole and spin of Kerr black hole (left panel). Degeneracy plot is for spin of Kerr BHs $a$ and the deviation parameter $\epsilon$ for the different values of the magnetic coupling parameter $\beta=0.25, 0, -0.25$. \label{iscobeta025}}
\end{figure*}

In Fig.\ref{iscobeta025} we provide the behavior of the ISCO radius of magnetic dipoles around quasi-Schwarzschild black hole in the presence and absence of external magnetic field (blue dashed, red dot-dashed and black solid lines in the top panel of the figure) and around rotating Kerr black hole in the absence of external magnetic field. One may see from the top panel of the figure that an increase of positive deviation parameter causes to decreases the ISCO radius while negative one increases. In some cases it is similar to the effect of spin of Kerr black hole. Moreover, positive (negative) values of magnetic coupling parameter shifts the ISCO radius outwards (towards) the central black hole. In the bottom panel we show (compare) effects of deviation parameter of quasi-Schwarzchild and spin of Kerr black holes for the degeneracy cases of the magnetic dipoles having the same values of ISCO radius. One may see from the degeneracy plot that negative deviation parameter can mimic the spin of Kerr black hole providing the same value of ISCO radius for corotating orbits of magnetized particles with the magnetic coupling parameter $\beta=0.25$  up to $a/M\simeq 0.3952$ while for the particles with the parameter $\beta=0$ and $\beta=-0.25$, it mimics up to the spin value of Kerr black hole $a\simeq0.1984M$ and $a\simeq 0.0537M$, respectively.  

Here, we will focus how the magnetic coupling parameter can mimic the spin of the Kerr black hole providing the same value for ISCO radius of magnetic dipoles around Schwarzschild black hole. One may easily calculate ISCO radius for magnetic dipoles around Schwarzschild black hole keeping the deviation parameter as zero.

\begin{figure*}[ht!]
	\begin{center}
		\includegraphics[width=0.48\linewidth]{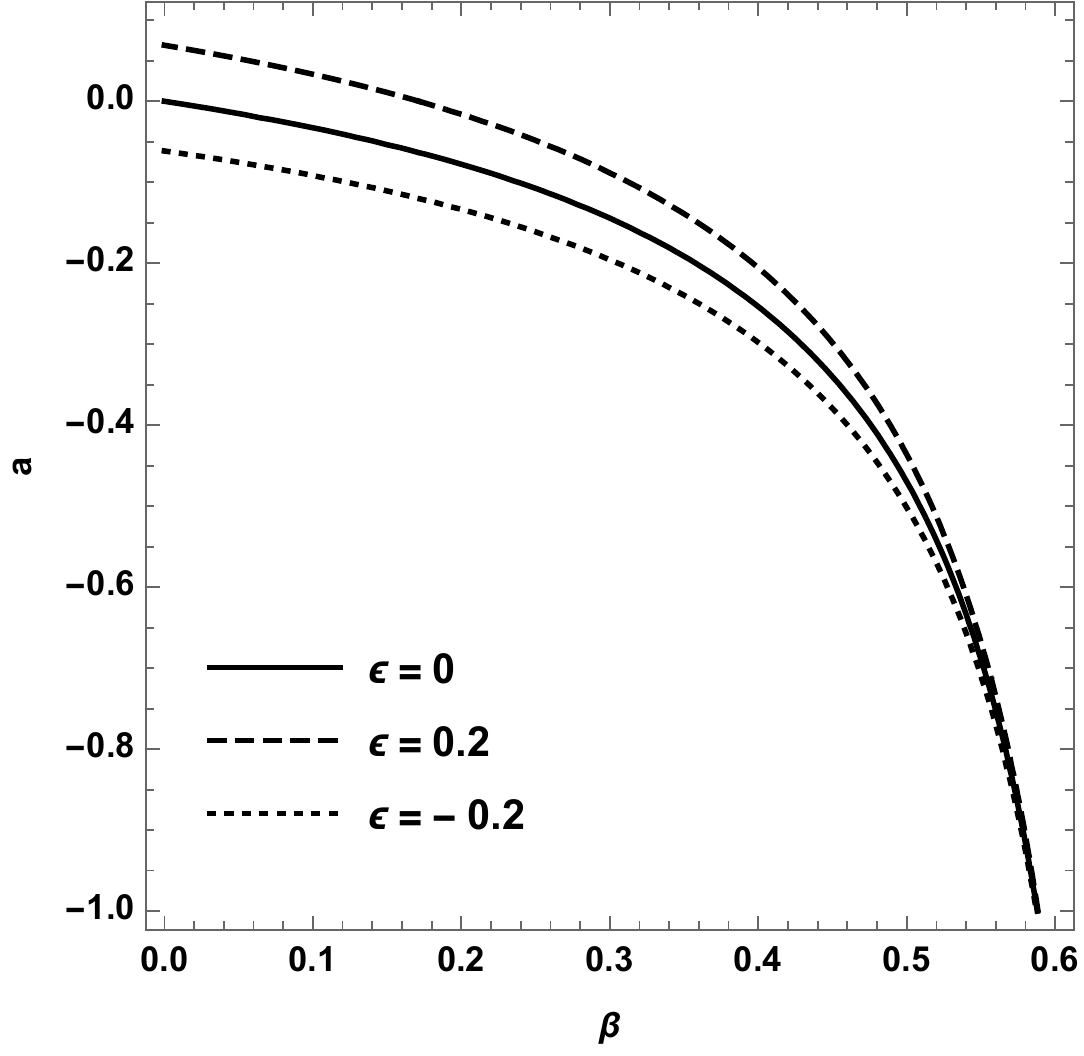}
		\includegraphics[width=0.48\linewidth]{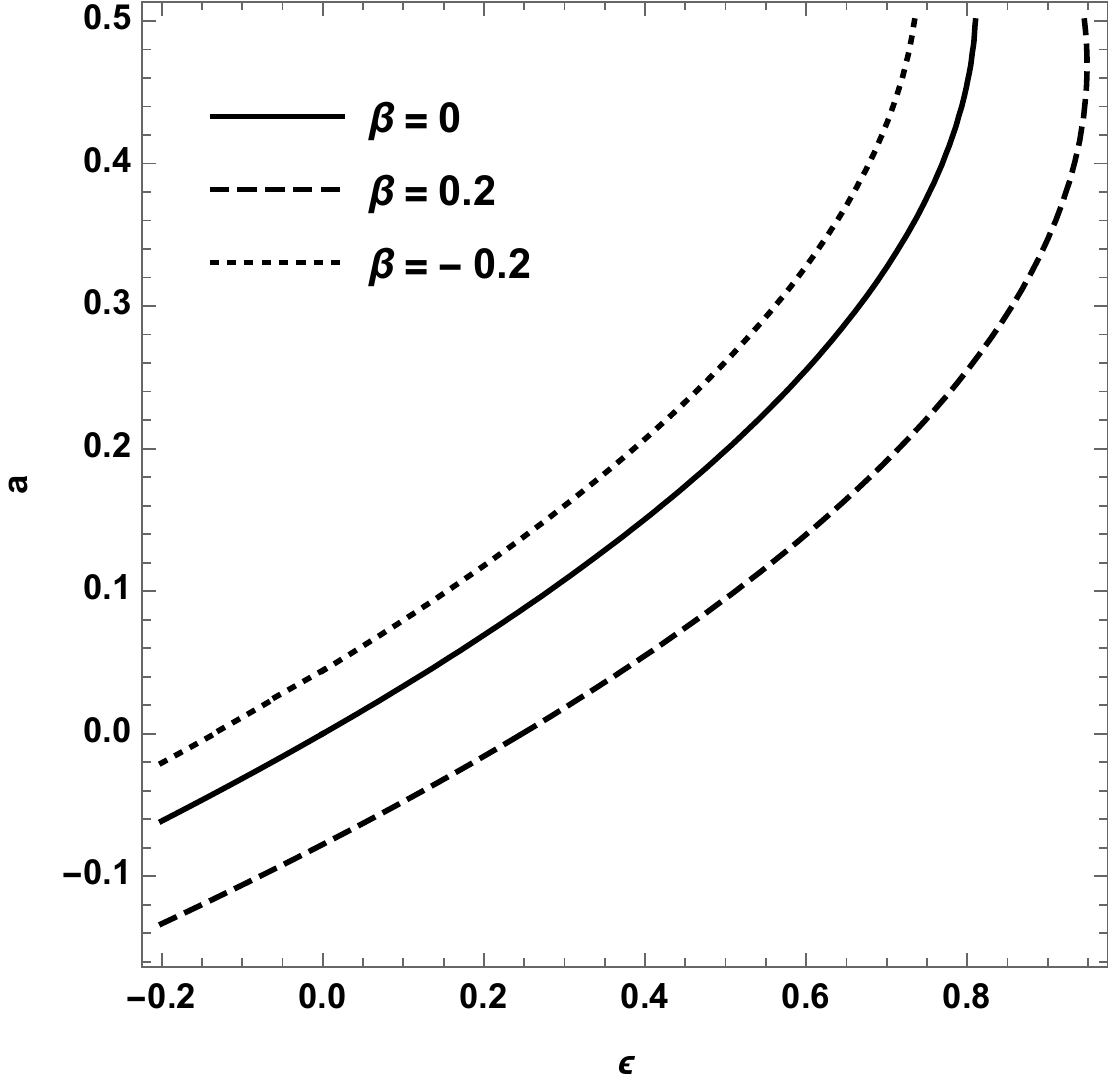}
	\end{center}
	\caption{Degeneracy plot between dimensionless spin of Kerr BHs $a$ and the magnetic coupling parameter $\beta$ and deviation parameter $\epsilon$ providing the same value for ISCO radius for magnetic dipoles around Schwarzschild black hole immersed in an external asymptotically uniform magnetic field. \label{ma_b}}
\end{figure*}

One can construct the degeneracy plot between magnetic parameter in quasi-Schwarzschild metric and rotation parameter of Kerr one. From the right panel of Fig.~\ref{misco} it is clearly seen that for fixed values of deviation parameter $\epsilon$ the ISCO radius goes up similar to the case when one increases the magnetic coupling parameter. It leads to conclusion that this coupling parameter should only mimic the rotation parameter of a Kerr metric for retrograde orbits. One can ensure this from the degeneracy plot as illustrated in Fig.~\ref{ma_b}

\begin{figure}[ht!]
	\begin{center}
		\includegraphics[width=0.9\linewidth]{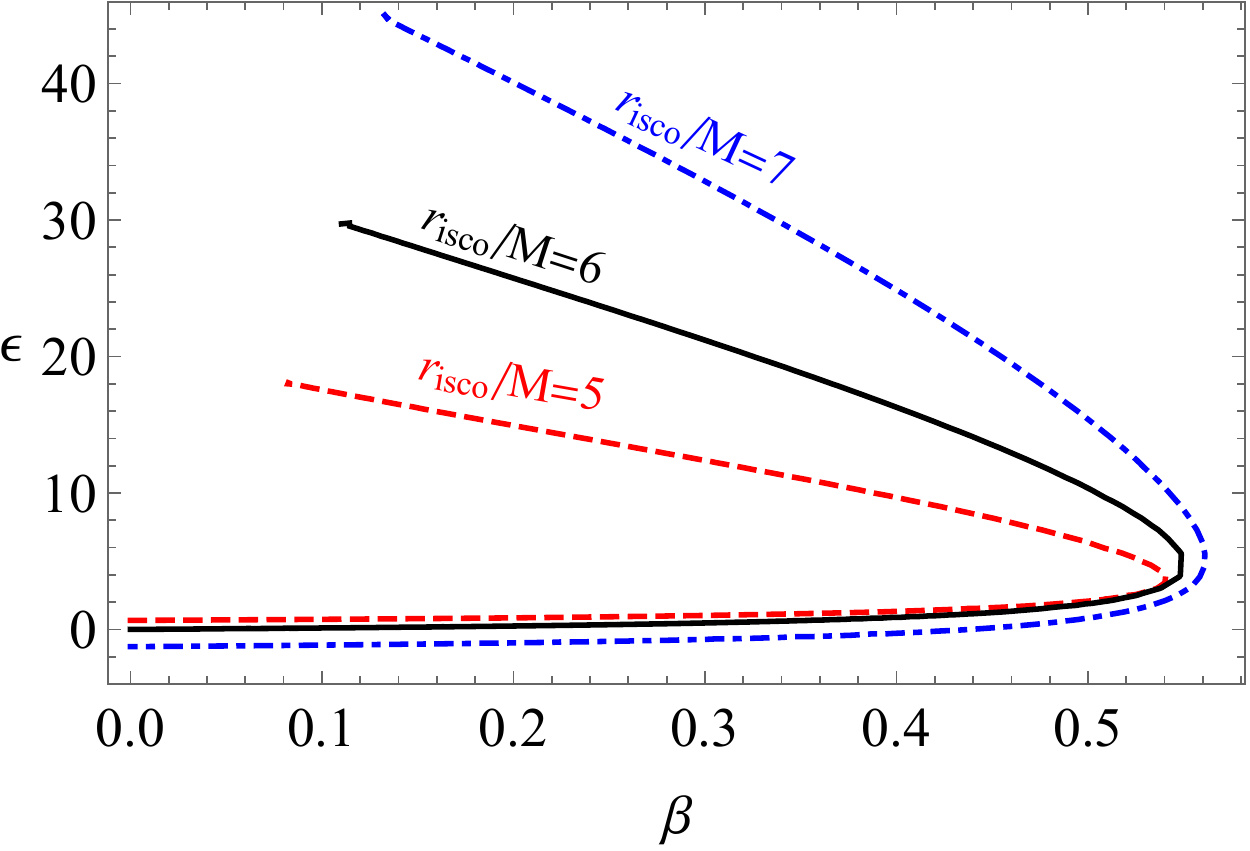}
	\end{center}
	\caption{Relations between the deviation parameter $\epsilon$ and magnetic coupling parameter $\beta$ for the fixed values of ISCO radius of magnetic dipoles.\label{me_b}}
\end{figure}

Degeneracy between values of parameter $\epsilon$ and magnetic coupling parameter for fixed values of ISCO radius is presented in Fig. ~\ref{me_b}. One can easily see that there are degeneracy values for the deviation parameter when the magnetic coupling parameter is fixed which  provide the same value for ISCO radius. Consequently, with the increase of ISCO radius the range of degeneracy values of the deviation parameter increases.  

\section{Conclusion \label{conclusion}}

In the work the motion of charged particles together with magnetic dipoles has been investigated to determine how well the spacetime deviation parameter $\epsilon$ and external uniform magnetic field can mimic rotation parameter of Kerr black hole which is the main point of this study.

Investigation of charged particle motion has shown that deviation parameter $\epsilon$ in the absence of external magnetic field can mimic the rotation parameter of Kerr spacetime up to $a\approx0.46$ which means that black hole assumed to be Kerr one with up to such rotation parameter can be also static quasi-Schwarzschild one with deviation parameter up to $\epsilon\approx0.8$. It has been also shown that the external magnetic field itself (i.e. without deviation parameter of spacetime) can mimic the rotation parameter up to $a\approx0.88$. The combination of these two parameters can do even better but not considerable job mimicking the rotation parameter up to $a>0.9$.
 
Study of the dynamics of magnetic dipoles around quasi-Schwarzschild black hole in the external magnetic field has shown that the maximum value of the effective potential for the fixed values of the specific angular momentum of the magnetic dipoles and deviation parameter of the spacetime around the BH increases with the increase of the magnetic coupling parameter and the positive deviation parameter increases the effective potential at the fixed values of the angular momentum and the magnetic coupling parameter while the negative ones decrease.  It is shown that there are degeneracy values of ISCO radius of test particles at $\epsilon_{cr}>\epsilon\geq 0.35$ which may lead to two different values of ISCO radius. Finally, we have studied how the deviation parameter mimics spin of Kerr BH providing the same values for ISCO radius of test particles. Since we consider magnetic dipoles as test ones, we have chosen here two different signs for the magnetic coupling parameter in the range of $\beta \in [-0.25,\ 0.25]$ and obtained that when the value of the deviation parameter in the range of $\epsilon \in (-1,\ 1)$ it can mimic the spin of rotating Kerr BH in the range $a/M \in (0.0537, \ 0.3952)$ for the magnetic dipoles with the values of magnetic coupling parameter $\beta \in [-0.25,\ 0.25]$ in corotating orbits. However, the mimic values of the spin parameter in counterrotating orbits lie in the range of $a/M \in (0.0152, \ 0.7863)$. We have pointed out that since the magnetic coupling parameter increases ISCO radius of the magnetic dipoles it can mimic the spin of Kerr BH only in corotating orbits up to its value $\beta=0.5922$. Moreover, we have shown that the degeneracy relations between magnetic coupling parameter and the deviation parameters for the fixed values of ISCO radius and found out that the ISCO radius can be the same at the two different positive values of the deviation parameter for the fixed values of magnetic coupling parameter. Performed study can be applied to the dynamics of magnetized matter and neutron stars in SMBH close environment.

\begin{acknowledgments}
	
This research is supported by the Uzbekistan Ministry for Innovative Development, Grants No. VA-FA-F-2-008 and No. MRB-AN-2019-29, the Innovation Program of the Shanghai Municipal Education Commission, Grant No.~2019-01-07-00-07-E00035, and the National Natural Science Foundation of China (NSFC), Grant No.~11973019. B.N. also acknowledges support from the China Scholarship Council (CSC), grant No.~2018DFH009013. This research is partially supported by an Erasmus+ exchange Grant between SU and NUUz.  A.A. is supported by a postdoc fund through PIFI of the Chinese Academy of Sciences.
	
\end{acknowledgments}

\bibliographystyle{apsrev4-1}  %% BibTeX style
\bibliography{gravreferences}

\end{document}